\documentclass[aps,prd,twocolumn,nofootinbib,showpacs,floatfix]{revtex4}
\usepackage{amsmath, amsthm}
\usepackage[dvips]{graphicx}

\newcommand{\lsim}{\lower0.6ex\vbox{\hbox{$ \buildrel{\textstyle <}\over{\sim}\ $}}}
\newcommand{\gsim}{\lower0.6ex\vbox{\hbox{$ \buildrel{\textstyle >}\over{\sim}\ $}}}
\newcommand{\beq}{\begin{equation}}
\newcommand{\eeq}{\end{equation}}

\newcommand{\Pktom}[2]{P_{\kappa}^{\mathrm{#1}\mathrm{#2}}}

\newcommand{\wlw}[1]{W_{\mathrm{#1}}}
\newcommand{\Da}{D_{\mathrm{A}}}

\newcommand{\ntomo}{N_{\mathrm{TOM}}}

\newcommand{\dd}{\mathrm{d}}

\begin{document}


\title{
Effects of Unstable Dark Matter on Large-Scale Structure and Constraints from Future Surveys
}

\author{Mei-Yu Wang and Andrew R. Zentner}
\affiliation{
Department of Physics and Astronomy \&
Pittsburgh Particle physics, Astrophysics, and Cosmology Center (PITT PACC), 
The University of Pittsburgh, 
Pittsburgh, PA 15260
}
\email{
mew56@pitt.edu,zentner@pitt.edu
}


\begin{abstract}

In this paper we explore the effect of decaying dark matter (DDM) on large-scale structure 
and possible constraints from galaxy imaging surveys.  DDM models have 
been studied, in part, as a way to address apparent discrepancies between the 
predictions of standard cold dark matter models and observations of galactic structure.  
Our study is aimed at developing independent constraints on these models.  
In such models, DDM decays into a less massive, stable dark matter (SDM) particle 
and a significantly lighter particle.  The small mass splitting between the parent DDM and the 
daughter SDM provides the SDM with a recoil or ``kick'' velocity $v_k$, inducing a 
free-streaming suppression of matter fluctuations.  
This suppression may be probed via weak lensing power spectra 
measured by a number of forthcoming imaging 
surveys that aim primarily to constrain dark energy.  
Using scales on which linear perturbation 
theory alone is valid (multipoles $\ell < 300$), 
surveys like Euclid or LSST can be sensitive to 
$v_k \gtrsim$ 90 km/s for lifetimes $\tau\sim 1-5$~Gyr.  
To estimate more aggressive constraints, we model nonlinear corrections 
to lensing power using a simple halo evolution model that is in good agreement 
with numerical simulations. In our most ambitious forecasts, using 
multipoles $\ell < 3000$, we find that imaging surveys can be sensitive to 
$v_k \sim 10$~km/s for lifetimes $\tau\lesssim 10$~Gyr.  Lensing 
will provide a particularly interesting complement to existing constraints 
in that they will probe the long lifetime regime ($\tau \gg H_0^{-1}$) far 
better than contemporary techniques.  A caveat to these 
ambitious forecasts is that the evolution of perturbations on nonlinear 
scales will need to be well calibrated by numerical simulations before they 
can be realized.  This work motivates the pursuit of such a numerical 
simulation campaign to constrain dark matter with cosmological weak lensing.

\end{abstract}

\date{\today}
\maketitle

\section{Introduction}
\label{section:introduction}

Many and various astronomical observations indicate that $\sim 5/6$ of the 
mass density of the Universe is non-baryonic {\em dark matter} 
(reviews include Refs.~\cite{jungman_etal96b,griest_kamionkowski00,bertone_etal05}). 
The simplest model of so-called {\em cold} dark matter (CDM) can be successfully 
applied to interpret an enormous amount of observational data, particularly those 
characterizing the large-scale ($ \gtrsim$~a few Mpc) structure of the Universe 
and the gross properties of galaxies.  In particular, 
the CDM model is consistent with the cosmic microwave background (CMB) anisotropy 
spectrum measured by the Wilkinson Microwave Anisotropy Probe (WMAP) and observations 
of the large-scale ($k \lesssim 0.1$~h/Mpc) galaxy clustering spectrum measured by the 
Sloan Digital Sky Survey (SDSS) \cite{Komatsu_etal11}.
On smaller scales, the situation is murkier.  Several observations indicate 
possible discrepancies between CDM theory and observations on smaller scales.  
Among these are the well-known {\em missing satellites problem} 
\cite{klypin_etal99b,moore_etal99} and the steep rotation curves of low-surface brightness 
galaxies \cite{deBlok_etal02,Simon_etal05,KuziodeNaray_etal08}.  
Exploring small-scale structure is challenging both 
observationally and theoretically.  On the theoretical 
side, it is necessary to model highly nonlinear phenomena to predict the properties 
of galaxies and the dark matter halos in which they reside.  Nevertheless, these 
potential shortcomings of CDM may point toward novel properties of dark matter and 
many alternatives to CDM have been considered, including {\em warm} dark matter (WDM) 
\citep{Colin_etal00,Bode_etal01,zentner_bullock02,zentner_bullock03,Lovell_etal11}
and self-interacting dark matter (SIDM) \citep{Spergel_etal00,peter_10,peter_etal10}.

As one alternative to CDM, unstable dark matter has been considered in a number of recent studies 
\cite{cen_01,Sanchez-Salcedo_03,Kaplinghat_05,Abdelqader_etal08,peter_10,peter_etal10,
peter_benson10,Bell_etal10,Bell_etal11,Aoyama_etal11}.
In such models, a dark matter particle of mass $M$ decays into 
a less massive daughter particle of mass $m=(1-f)\,M$ and a 
significantly lighter, relativistic particle, with a lifetime on 
the order of the age of the Universe.  In decaying dark matter models, 
cosmological structure growth is altered in a time- and scale-dependent manner 
\cite{Kaplinghat_05,wang_zentner10}.  We explored models with $f \simeq 1$ 
in a previous paper \cite{wang_zentner10}.  
If $f \ll 1$, the stable dark matter (SDM) daughter particle will receive 
a non-relativistic {\em kick} velocity, $v_k \simeq f\,c$.  This kick can 
be sufficient to alter small-scale structure growth, modifying the 
predicted structures and abundances of the dark matter halos that 
host galaxies.  Previous studies have explored just this possibility and 
placed limits on these scenarios by demanding that the alterations to 
structure growth not be so severe as to destroy the successes that 
CDM theory enjoys on large scales.  
An upper bound on the lifetime of the decaying dark matter (DDM) 
particle of $\tau \gtrsim 30-40$~Gyr for 
$v_k \gtrsim 100$~km/s can be derived from the observed galaxy-cluster mass function 
and galaxy mass-concentration relation \cite{peter_10,peter_etal10}.  Lifetimes 
$\tau \lesssim 30$~Gyr with $20\ \mathrm{km/s} \lesssim v_k \lesssim 200\ \mathrm{km/s}$ 
may be in tension with the observed properties of the Milky Way satellite galaxy population, 
but uncertainties associated with the details of nonlinear structure growth in these models 
as well as the star formation histories of Local Group galaxies are significant \cite{peter_benson10}.
Significantly tighter constraints can be obtained if the daughter particles are 
standard model particles \cite{Bell_etal10}.

With a number of forthcoming large-scale galaxy surveys being undertaken this decade, such 
as the Dark Energy Survey (DES)\footnote{{\tt http://www.darkenergysurvey.org}},
the Panoramic Survey Telescope and Rapid Response System 
(Pan STARRS) \footnote{{http://pan-starrs.ifa.hawaii.edu/}}, the Large Synoptic Survey Telescope (LSST)\footnote{{\tt http://www.lsst.org}} \cite{lsstbook},
Euclid\footnote{{\tt http://sci.esa.int/euclid}} \cite{eicbook}, and WFIRST\footnote{{http://wfirst.gsfc.nasa.gov}}, it will 
be possible to investigate possible subtle imprints of decaying dark matter on the relatively 
large-scale structure of the Universe.  In DDM models, the kick velocity at decay 
imparts upon the stable daughter particles the ability to smooth gravitational 
potential perturbations on scales smaller than the classic free-streaming scale. 
This behavior is similar to the cosmological influence of massive neutrinos or 
WDM, and numerous studies have shown that these features may be detectable in 
large-scale structure through galaxy clustering \citep{Reid_etal10}, Lyman-$\alpha$ 
forest \citep{Seljak_etal06,boyarsky_etal08,Viel_etal10}, and cosmological weak lensing 
\citep{Hannestad_etal06,Ichiki_etal09,Markovic_etal11}.

In this paper, we explore the effect of DDM model on large-scale structure 
and study the possible independent constraints on these models from forthcoming 
weak lensing surveys. 
There are distinct advantages to this approach.  First, DDM models are explored largely 
in order to mitigate the possible small-scale problems of CDM, so it is necessary to 
explore independent predictions and probes in order to test such models.  Second, 
the effects of DDM on large-scale structure, and on weak lensing power spectra in particular, 
are simpler to model in that they do not require detailed knowledge of galaxy formation 
in the highly nonlinear regime including star formation histories, star formation and 
active galaxy feedback, scale-dependent galaxy clustering bias and numerous other complicated 
phenomena that are known to be important on small scales.  Third, many surveys already have the 
goal of measuring cosmological weak lensing as a probe of dark energy \cite{detf}, so this 
test can be performed largely with the observational infrastructure used to study 
dark energy at no additional cost. 
Lyman-$\alpha$ forest spectra provide an additional, 
promising method to constrain DDM, but also introduce complications associated with 
modeling the clustering of neutral hydrogen along lines of sight to high redshift.  
We will present Lyman-$\alpha$ constraints on DDM, and forecasts for future 
measurements of the Lyman-$\alpha$ forest, in a forthcoming follow-up paper.

In the following sections, 
we will show that the imprint of DDM on lensing power spectra is sufficiently distinct 
from other cosmological parameters, such as neutrino mass, that one can disentangle 
degeneracies among them.  Indeed, forthcoming data will be able to disentangle the 
two parameters of DDM models, lifetime and kick velocity (or equivalently $f$).   
Finally, large-scale lensing surveys will provide, at minimum, competitive and 
independent constraints on DDM models exploiting only scales on which linear 
cosmological perturbation theory is appropriate.  These probes will have the 
particular advantage of probing significantly larger DDM lifetimes than small-scale 
structure studies and will not require detailed modeling of galaxy 
formation.  Moreover, a relatively modest numerical simulation program may 
enable one to use mildly nonlinear scales $k \lesssim 1$~h/Mpc to obtain constraints 
on DDM that exceed current constraints and are robust to uncertainties associated 
with star formation and feedback in small galaxies.

The outline of this paper is as follows. 
In \S~\ref{section:Weak Lensing}, we describe cosmological weak lensing observables.  
In \S~\ref{section:models}, we discuss the perturbation evolution of both the parent 
and daughter particles in DDM models. In \S~\ref{section:nonlinear}, we study the effect 
of DDM on dark matter halo density profiles and explore possible influences of DDM on nonlinear 
structure. In \S~\ref{section:fisher}, we describe the methods we use to compute constraints 
on DDM model parameters. We present our results in \S~\ref{section:results}, beginning with 
the general effects of DDM on large-scale structure, including a detailed discussion of 
the free-streaming of the daughter SDM particles. We also give our forecasts for 
constraints on DDM from large-scale weak lensing surveys and compare possible future 
bounds with existing limits. In \S~\ref{section:conclusion} we summarize our work.

\section{Weak Gravitational Lensing Observables}
\label{section:Weak Lensing}

Weak lensing as a cosmological probe 
has been discussed at length in numerous papers 
(a recent review is Ref.~\cite{huterer10}).  We give a brief 
description of our methods below, which are based on the 
conventions and notation in Ref.~\cite{zentner_etal08}. The most robust 
forecasts derive from considerations of possible weak lensing 
measurements restricted to scales where linear perturbative     
evolution of the metric potentials remains useful. However, we attempt to 
estimate possible improvements to the constraining power 
of weak lensing observables, provided that mildly nonlinear evolution can be 
modeled robustly.

We consider the set of observables that may be available from 
large-scale galaxy imaging surveys to be the auto- and 
cross-spectra of lensing convergence from sets of galaxies in 
$\ntomo$ redshift bins.  The $\ntomo (\ntomo+1)/2$ distinct 
convergence spectra are
\beq
\label{eq:pkij}
\Pktom{i}{j}(\ell) =\int \dd z  \frac{\wlw{i}(z)\wlw{j}(z)}{H(z) D_A^2(z)}P_m(k=\ell/\Da,z),
\eeq
where $i$ and $j$ label the redshift bins of the source galaxies.  We take 
$\ntomo=5$ and evenly space bins in redshift from a minimum redshift 
of $z=0$ to a maximum redshift of $z=3$. Increasing the number of bins 
beyond $\ntomo=5$ adds only negligibly to the constraining power of 
lensing data, in accord with an analogous statement for dark energy 
constraints \cite{ma_etal06}.

In Eq.~(\ref{eq:pkij}), 
$H(z)$ is the Hubble expansion rate, 
$D_A(z)$ is the comoving angular diameter distance, 
and  $P_m(k,z)$ is the matter power spectrum at wavenumber $k$ and redshift $z$.  
In the following section, we describe our use of the publicly-available {\tt CMBFAST} code to calculate
$P_m(k)$ from cosmological perturbation evolution. In this case it will be more natural 
to work in the synchronous gauge, and transforming between different gauge systems
can be accomplished straightforwardly by 
following, for example, the methods described in Ref.~\citep{ma_bertschinger95}.

The $W_{i}$ are the lensing weight functions for source galaxies 
in redshift bin $i$.  In practice, the galaxies will be binned by 
photometric redshift, so that the bins will have non-trivial overlap 
in true redshift (see Refs.~\cite{ma_etal06,hearin_etal10} for detailed discussions).  
Defining the true redshift distribution of source galaxies in the 
$i$th photometric redshift bin as $dn_i/dz$, the weights are 
\beq
\label{eq:W}
W_{i}(z)={3 \over 2} \Omega_M H_0^2(1+z)D_A(z) \int dz' {D_A(z,z') \over D_A(z')} {dn_i \over dz'}
\eeq
where $D_A(z,z')$ is the angular diameter 
distance between redshift $z$ and $z'$ and $H_0$ is the present Hubble rate.

We model the uncertainty induced by utilizing photometric galaxy redshifts 
with the probability function of assigning an individual source galaxy photometric 
redshift $z_p$ given a true redshift $z$, $P(z_p|z)$.  The true 
redshift distribution of sources in the $i$th photometric redshift bin is
\beq
\label{eq:dnidz}
\frac{dn_i(z)}{dz} = \int^{z_{p,i}^{(high)}}_{z_{p,i}^{(low)}}\ dz_p\ \frac{dn(z)}{dz}\ P(z_p|z)
\eeq
Here we take the true redshift distribution to be 
\beq
\label{eq:dndz}
{dn(z) \over dz}=\bar{n} {4z^2 \over \sqrt{2 \pi z_0^3}} \exp[-(z/z_0)^2]
\eeq
with $z_0 \simeq 0.92$, so that the median survey redshift to $z_{med}$ = 1, 
and $\bar{n}$ as the total density of source galaxies per
unit solid angle \cite{smail_etal95a,smail_etal95b,newman08}.  
We assume that uncertain photometric redshifts can be approximated 
by taking
\beq
\label{eq:p}
P(z_p|z)={1 \over \sqrt{2 \pi \sigma_z}} \exp\left[-{(z_{p}-z)^2 \over 2\sigma_z^2}\right]
\eeq
where $\sigma_z(z)=0.05(1+z)$ \citep{ma_etal06}. Complexity 
in photometric redshift distributions is an issue that must 
be overcome to bring weak lensing constraints on cosmology 
to fruition (e.g., Ref.~\cite{bernstein_huterer10,hearin_etal10}).  

Observed convergence power spectra $\bar{P}^{ij}_{\kappa}(\ell)$, 
contain both signal and shot noise,
\beq
\label{trueP}
\bar{P}^{ij}_{\kappa}(\ell)=P_{\kappa}^{ij}+n_{i}\delta_{ij}\langle \gamma^2 \rangle 
\eeq
where $\langle \gamma^2 \rangle$ is the noise from 
intrinsic ellipticities of source galaxies, 
and $n_i$ is the surface density of galaxies in the 
ith tomographic bin.  We follow recent convention 
and set $\sqrt{\langle \gamma^2 \rangle}=0.2$, subsuming 
additional errors on galaxy shape measurements into an 
effective mean number density of galaxies, $\bar{n}$.  
Assessments of intrinsic shape noise per galaxy may 
be found in, for example \cite{massey_etal04,kasliwal_etal08,lsstbook}.  
Assuming Gaussianity of the lensing field, the covariance between 
observables $\bar{P}^{ij}_{\kappa}$ and $\bar{P}^{kl}_{\kappa}$ is
\beq
\label{cab}
C_{AB}=\bar{P}^{ik}_{\kappa}\bar{P}^{jl}_{\kappa}+\bar{P}^{il}_{\kappa}\bar{P}^{jk}_{\kappa}
\eeq
where the i and j map to the observable index A, and k and l map to B such that 
$C_{AB}$ is a square covariance matrix with $\ntomo (\ntomo+1)/2$ rows and columns.  
We assume Gaussianity throughout this work and even in our most 
aggressive forecases we consider only multipoles 
$\ell < 3000$, at which point the Gaussian assumption and several weak lensing 
approximations break down 
\cite{white_hu00,cooray_hu01,vale_white03,dodelson_etal06,semboloni_etal06}.

\section{Decaying Dark Matter Models}
\label{section:models}

\subsection{Evolution of the Average Properties of Unstable Dark Matter}
\label{subsection:collision}

We consider dark matter decays into another species of stable 
dark matter with a small mass splitting, DDM~$\rightarrow$~SDM~$+$~L, 
where L denotes a ``massless'' daughter particle, SDM is the stable dark matter with mass $m$, 
and DDM is the decaying dark matter with mass $M$. 
The mass loss fraction f of DDM is directly related to the kick velocity 
deposited to the SDM particle by $f \simeq v_k/c$ from energy-momentum conservation.  
The following relations are valid in the rest frame of DDM particles with 
the kick velocity of SDM being the velocity relative to the DDM rest frame.

For a general decay, neglecting Pauli-blocking factors and inverse decays, 
the rate of change in the DDM distribution function is
\begin{equation}
\label{eq1}
\dot{f}_{DDM}(q_{DDM})=-{aM\Gamma \over E_{DDM}} f_{DDM}(q_{DDM}), 
\end{equation}
where $\dot{f}$ denotes the partial derivative of the distribution 
function with respect to conformal time, $\dd \tau = \dd t/a$, 
$\Gamma$ is the decay rate, $a$ is the cosmological 
scale factor, $q_{DDM}$ is the comoving momentum, and $E_{DDM} = \sqrt{q_{DDM}^2 + M^2 a^2}$.
Specializing to two-body decays, one can show that
the corresponding change to the SDM distribution 
function will be \cite{Starkman_etal94,Kaplinghat_05}
\begin{equation}
\label{eq2}
\dot{f}_{SDM}(p_{SDM})={aM^2 \Gamma \over 2 E_{SDM}\,p_{SDM}\, p_{CM}} \int^{E_f}_{E_i\,} dE\, f_{DDM}(p) ,
\end{equation}
where $$E_{f,i} = {\frac{1}{2}E_{SDM}m_0^2}\pm p_{SDM}p_{CM}M/m^2_{SDM},$$ 
the quantity $p_{CM}$ is the center-of-mass momentum, and $m_0^2 \equiv M^2+m^2$.

We define the average distribution function, $f_{i}^0(q,\tau)$, 
and the perturbation to the distribution function, $\Psi_{i}(\vec{x},\vec{q},\tau)$, 
for each species of particle labeled by $i$, according to
\begin{equation}
\label{eq7}
f_{i}(\vec{x}, \vec{q}, \tau) =f_{i}^0(q, \tau)[1+\Psi_i(\vec{x}, \vec{q}, \tau)]
\end{equation}
where $i$ can be the DDM, SDM, or L.  
Since DDM particles are non-relativistic, their zero order 
phase-space distribution is the Maxwell-Boltzmann function.
The zero order phase-space distribution 
function of SDM is \cite{Aoyama_etal11,Kaplinghat_05}
\begin{equation}
\label{eq17}
f_{0,SDM}(q,a)={\Gamma \Omega_M \rho_{crit} \over M q^3 H(a')} \exp(-\Gamma t_q)\, \Theta(a p_{CM}-q)
\end{equation}
where $q$ is the comoving momentum of the SDM particle, 
$a'=q/p_{CM}$, and $t_q=t(a')$. This can be derived from the fact that 
the decay always generates SDM particles with the same physical momentum 
$p_{CM}$. In the SDM distribution function, the spectrum of different momenta
arises from decays at different times, designated by the cosmic scale factor 
$a'$ so that $q=p_{CM}a'$.  
The Heaviside step function $ \Theta(a p_{CM}-q)$ (see Eq.~\ref{eq17}) enforces a 
cut-off $q_{\rm max}=ap_{CM}$ at a given redshift $a$. This maximum momentum stems from 
the fact that the maximum momentum at a given redshift is from decay processes 
happening at that time, while SDM with lower momenta are from the earlier decays. 
To be explicit, the average comoving number density of SDM particles 
is the integral of $f_0$ over momentum space, 
\begin{align}
\label{eq18}
& n_{SDM}=\int q^2 \dd q\, \dd \Omega\, f_{0,SDM}(q) \\
& \to \dd n_{SDM}(q) =4\pi q^2 \dd q \, f_{0,SDM}(q) 
\end{align}
Thus $f_{0,SDM}$ can be written as
\begin{align}
\label{eq17.5}
&f_{0,SDM}(q)={\dd n_{SDM}(q) \over q^2 \dd q} ={ \dd n_{SDM} \over q^2 p_{CM} \dd a'}
={1 \over H(a') q^3}{\dd n_{SDM} \over \dd t'} \\
& \to f_{0,SDM}(q)={1 \over MH(a') q^3}{\dd (\rho_{DDM} a'^3) \over \dd t'}
\end{align}
This then implies Eq.~(\ref{eq17}) after enforcing the maximum momentum at $q_{\rm max}=ap_{CM}$. 

The evolution equations for the mean energy densities in the two dark matter components are given
by the integrals of Eq.~\ref{eq1} and Eq.~\ref{eq2} using the unperturbed distribution 
function.  They read
\begin{equation}
\label{eq3}
\dot{\rho}_{DDM}+3 {\dot{a}\over a} \rho_{DDM}=-a \Gamma \rho_{DDM}
\end{equation}
and
\begin{equation}
\label{eq4}
\dot{\rho}_{SDM}+3 {\dot{a}\over a} (\rho_{SDM}+P_{SDM})= \Gamma {a m_0^2 \over 2M^2} \rho_{DDM}
\end{equation}
respectively.  Given the DDM energy density, the decay product 
energy density $\rho_d=\rho_{SDM}+\rho_{L}$ can be obtained using the first law of 
thermodynamics \cite{Lopez_etal99,Kolb_Turner} from
\begin{equation}
\label{eq5}
{\dd {a^3 \rho_d}\over \dd \tau} =-P_d {\dd a^3 \over \dd \tau} - {\dd (a^3 \rho_{DDM})\over \dd \tau}.
\end{equation}
This implies that the energy density evolution of the massless daughter particle L is
\begin{equation}
\label{eq6}
\dot{\rho}_{L}+3 {\dot{a}\over a} (\rho_{L}+P_{L})=\dot{\rho}_{L}+4 {\dot{a}\over a} \rho_{L}=\Gamma {a (M^2-m^2) \over 2M^2} \rho_{DDM}
\end{equation}

\subsection{Perturbations}
\label{subsection:perturbation}

To compute the contemporary lensing power spectra, it is necessary to compute the perturbations 
to the dark matter distributions and the metric.  Our treatment of perturbations follows 
the conventions established in \citet{Ma_etal95}.  We will present our results in the synchronous gauge, 
because this choice lends itself to numerical evaluation.  In the synchronous gauge, the Fourier 
transform of the Boltzmann equation can be written 
\begin{equation}
\label{eq7.5}
{\partial \Psi\over \partial \tau} +i {q\over E} (\vec{k} \cdot \hat{n}) \Psi 
+ {\dd \ln f_0\over \dd \ln q} \left[ \dot{\eta} - {\dot{h}+6\dot{\eta}\over 2} (\hat{k} \cdot \hat{n})^2 \right] 
= {1 \over f_0} \left( {\partial f \over \partial \tau}\right)_C
\end{equation}
The DDM perturbation equations are the same as the well-known equations 
describing CDM (see Ref.~\cite{Ma_etal95}), so we will not describe them 
any further. The term on the right-hand side of Eq.~(\ref{eq7.5}) is the 
so-called {\em collisional term} representing the change in the distribution 
functions due to interactions (decays in our case).  In the absence of 
non-gravitational interactions, $( \partial f/\partial \tau )_C = 0$.  For the 
decay products, the collision terms are
\begin{equation}
\label{eqc3}
\left( {\partial f_{SDM} \over \partial \tau}\right)_C={a m_0^2 \Gamma \over 2 M E} f_{0,DDM} (1+\Psi_{DDM})
\end{equation}
and
\begin{equation}
\label{eqc4}
\left( {\partial f_{L} \over \partial \tau}\right)_C={a (M^2-m^2)\Gamma \over 2 M E} f_{0,DDM} (1+\Psi_{DDM}).
\end{equation}
The factors $m_0^2/(2M^2)$ and $(M^2-m^2)/(2M^2)$ that appear in the 
SDM and L collision terms can be easily understood.  Consider a two-body decay 
in the rest frame of the DDM particle, $A \rightarrow B+C$, with corresponding masses 
$m_A$, $m_B$, and $m_C$.  The energies of $B$ and $C$ in the rest frame of $A$ 
are $E_B=(m_A^2+m_B^2-m_C^2)/(2 m_A)$ and $E_C=(m_A^2+m_C^2-m_B^2)/(2 m_A)$. 
So these factors represent the ratios of energy that have been deposited into different daughter particle species.

The perturbations for the massless relativistic daughter particles may be treated in 
a manner analogous to that of massless neutrinos, save for the peculiar 
distribution function of the L.  
We integrate out the momentum dependence in the distribution function by defining
(in Fourier space)
\begin{equation}
\label{eq8}
F_{L}(\vec{k}, \hat{n}, \tau)={\int q^2 \dd q\, q f_{L}^0 (q,\tau)\, \Psi_{L}(\vec{k}, q, \hat{n},\tau) \over \int q^2 \dd q\, q f_{L}^0 (q,\tau)}
\end{equation}
An expansion of $F_{L}$ in a series of Legendre polynomials $P_{l}(\hat{k} \cdot \hat{n})$
has the form
\begin{equation}
\label{legendre}
F_{L}(\vec{k}, \hat{n}, \tau)=\Sigma^{\infty}_{l=0} (-i)^l (2l+1)F_{L,l}(\vec{k}, \tau)\, P_l(\hat{k} \cdot \hat{n}).
\end{equation}
The standard synchronous gauge perturbations in density, velocity, and anisotropic stress are
\begin{equation}
\label{deltadef}
\delta_{\rm L} = F_{L,0},
\end{equation}
\begin{equation}
\label{thetadef}
\theta_{\rm L} = \frac{3}{4}k\, F_{L,1},
\end{equation}
and
\begin{equation}
\label{sigmadef}
\sigma_{\rm L} = \frac{1}{2} F_{L,2}
\end{equation}
respectively.  Evaluating the Boltzmann equation for our Legendre polynomial expansion in 
Eq.~(\ref{legendre}) yields the evolution of the multipole coefficients in the conventional 
notation, 
\begin{equation}
\label{eq11}
\dot{\delta}_{\rm L}=-{2 \over 3}(\dot{h}+2\theta_{\rm L})+a \Gamma {E_2 \over M}{\rho_{\rm DDM} \over \rho_{\rm L}}(\delta_{\rm DDM}-\delta_{\rm L}),
\end{equation}
\begin{equation}
\label{eq12} 
\dot{\theta}_{\rm L}=k^2({\delta_{\rm L} \over 4}-\sigma_{\rm L})-a \Gamma {E_2 \over M} {\rho_{\rm DDM} \over \rho_{\rm L}}\theta_{\rm L},
\end{equation}
\begin{equation}
\label{eq13} 
\dot{\sigma}_{\rm L}={2 \over 15}(2\theta_{\rm L}+\dot{h}+6\dot{\eta}-{9 \over 4}kF_{\rm L,3})-a \Gamma{E_2 \over M} {\rho_{\rm DDM} \over \rho_{\rm L}}\sigma_{\rm L},
\end{equation}
and
\begin{equation}
\label{eq14} 
\dot{F}_{\rm L, l}={k \over 2l+1}[lF_{{\rm L},l-1}-(l+1)F_{{\rm L},l+1}]-a \Gamma {E_2 \over M}{\rho_{\rm DDM} \over \rho_{\rm L}}F_{{\rm L},l} ,\; l\geq 3,
\end{equation}  
Here we have defined $E_1=(M^2+m^2)/(2M)=m_0^2/(2M)$ and $E_2=(M^2-m^2)/(2M)$.

The SDM must be treated differently to account for their finite mass and non-trivial velocity kicks.  
We expand the perturbation to the distribution function, $\Psi$, in a Legendre series
\begin{equation}
\label{eq19} 
\Psi(\vec{k}, \hat{n}, q, \tau) =\sum^\infty_{l=0} (-\imath)^l (2l+1)\Psi_l (\vec{k}, q, \tau) P_l(\hat{k} \cdot \hat{n}).  
\end{equation}  
We have dropped the ``SDM'' subscript on $\Psi$ for brevity as there should be no cause for confusion in 
this context.  Evaluating the Boltzmann evolution equation on this expansion, we obtain for the different 
multipoles
\begin{multline}
\label{eq21} 
{\partial {\Psi}_0 \over \partial \tau}=-{qk \over E} \Psi_1 + {1\over 6} \dot{h} {d\ln f_{SDM, 0} \over d \ln q} +a \Gamma {E_1 \over E} {f_{DDM,0} \over f_{SDM,0}} \Psi_{DDM,0}\\
-a \Gamma {E_1 \over E} {f_{DDM,0}\over f_{SDM,0}}{\Psi}_0 ,
\end{multline}
\begin{equation}
\label{eq22} 
{\partial {\Psi}_1 \over \partial \tau}={qk \over 3 E} (\Psi_0 -2\Psi_2)-a\Gamma {E_1 \over M} {f_{DDM,0}\over f_{SDM,0}}{\Psi}_1 ,
\end{equation}
\begin{multline}
\label{eq23} 
{\partial {\Psi}_2 \over \partial \tau}={qk \over 5 E} (2\Psi_1 -3\Psi_3)-({1\over 15}\dot{h}+{2\over 5}\dot{\eta}){d f_{SDM, 0} \over d ln q}\\
-a\Gamma {E_1 \over E} {f_{DDM,0}\over f_{SDM,0}}{\Psi}_2 ,
\end{multline}
and
\begin{equation}
\label{eq24} 
{\partial {\Psi}_l \over \partial \tau}={qk \over (2l+1) E} (l \Psi_{l-1} -(l+1)\Psi_{(l+1)})-a\Gamma {E_1 \over M} {f_{DDM,0}\over f_{SDM,0}}{\Psi}_l.
\end{equation}  
for $l\geq$ 3.

If we restrict attention only to cases in which the mass difference between the DDM and SDM particles is small, 
$f=1-m/M \ll 1$, the SDM particle will receive an 
extremely non-relativistic kick velocity $v_k\simeq f c$. As we should expect, 
SDM behaves similarly to CDM, aside from the fact that it is endowed with a 
non-negligible distribution of momentum due to the DDM decays.  In this limit, the SDM 
perturbations evolve as for a standard non-relativistic dark matter species, 
\begin{equation}
\label{eq27} 
\dot{\delta}_{SDM} = -\theta_{SDM}-{1 \over 2} \dot{h} +a \Gamma {E_1 \over M} {\rho_{DDM} \over \rho_{SDM}}(\delta_{DDM}-\delta_{SDM})
\end{equation}
and
\begin{equation}
\label{eq28} 
\dot{\theta}_{SDM} = -{\dot{a} \over a}\theta_{SDM}+{\delta P_{SDM} \over \delta \rho_{SDM}}k^2 \delta_{SDM}-a \Gamma {E_1 \over M} {\rho_{DDM} \over \rho_{SDM}}\theta_{SDM},
\end{equation}
where 
\begin{equation}
\label{eq29} 
c_s^2={\delta P_{SDM} \over \delta \rho_{SDM}}={{ 4 \pi \over 3} a^{-4} \int q^2 dq {q^2 \over E} f_0(q) \Psi_0 \over { 4 \pi} a^{-4} \int q^2 dq {E} f_0(q) \Psi_0}
\end{equation}
The higher multipole terms become negligible in the non-relativistic as they 
are proportional to powers of the ratio of the kinetic energy to the 
total energy, $q/\epsilon$.

Though we solve the complete equations for the evolution of the SDM perturbations, the non-relativistic kick velocity 
approximation is valid in most of our calculations.  The most interesting constraints from future surveys are relevant 
for models with $v_k \lesssim 10^{-3} c$ and relativistic kicks have already been ruled out for a wide range of 
lifetimes \cite{peter10,peter_benson10}.


Perturbation growth is suppressed on scales smaller than the {\em free-streaming} scale.  The 
free-streaming scale is, in turn, determined by an integral of the sound speed $c_s$.  We defer 
a detailed discussion of the free-streaming scale in our decaying dark matter models and its 
imprint on the matter and lensing power spectra to \S~\ref{section:results}.

\section{Nonlinear Corrections to Structure Growth}
\label{section:nonlinear}

Our most robust constraints stem from perturbations on linear scales.  However, 
it is interesting to estimate the level of constraints that may be achieved by 
exploiting mildly nonlinear scales as is common practice in the established 
framework for exploring dark energy with lensing and galaxy clustering statistics \cite{detf}.  
Including mildly nonlinear scales improves constraints because it increases the 
signal-to-noise of lensing measurements and because it includes information 
regarding the effects of DDM on the abundance and internal structures of cluster-sized 
dark matter halos.  We explore constraints including mildly nonlinear scales as a means 
of estimating the level of constraints that may be achievable after an exhaustive 
numerical simulation program, similar to what is being performed for dark energy \cite{heitmann_etal08}.

We implement the nonlinear corrections to the matter and lensing power spectra using the 
halo model \cite{cooray_sheth02}.  The halo model is known to exhibit mild systematic 
offsets compared to numerical simulations and the nonlinear correction of Ref.~\cite{smith_etal03}.  
However, we use the halo model because it provides a convenient framework for estimating the 
alterations to nonlinear structure induced by DDM before performing an exhaustive numerical investigation.  
We combine the standard aspects of the halo model with an analytical model proposed by 
\citet{Sanchez-Salcedo_03} for the alterations to dark matter halo structure 
due to the kick velocities generated in the decay process.

\begin{figure}[ht!]
\includegraphics[height=9.2cm]{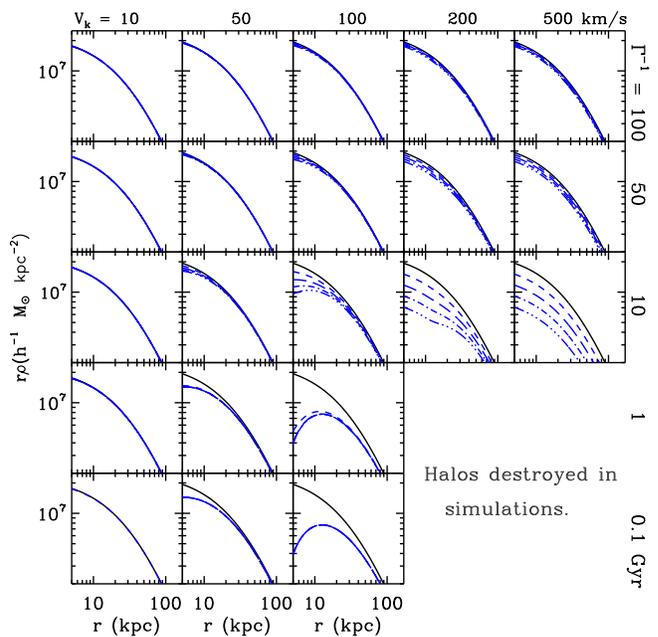}
\caption{ 
Dark matter density profiles times radius, $r\,\rho(r)$, 
as a function of radius and time. 
The dark matter halo mass is $M_{\rm h}=10^{12}\, \mathrm{M}_\odot$.  In the absence of 
dark matter decays, the halo concentration is $c=5$. The halo has a virial
speed $v_{vir} \equiv \sqrt{GM_{\rm h}/R_{vir}} \approx$~130 km/s. 
Different panels are for different choices of kick velocity and lifetime as labeled 
along the top and right axes respectively. In each
panel the solid lines show the initial NFW profile. The short-dashed line, 
long-dashed line, dash-dotted line, and dash-double-doted line represent
density profiles after 2.5, 5, 7.5, and 10 Gyr. 
This figure is designed to be directly comparable to the simulation 
results displayed in Fig. 1 of Ref.~\cite{peter_etal10}.
}
\label{fig:den_peter}
\end{figure}

\begin{figure}[ht!]
\includegraphics[height=9.2cm]{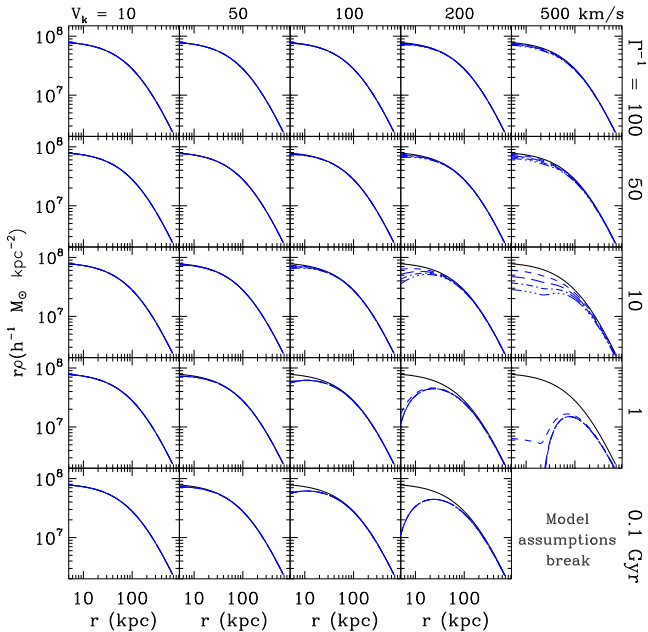}
\caption{ 
Similar to Figure~\ref{fig:den_peter} but for halos with $M_{\rm h} = 5 \times 10^{13}\, \mathrm{M}_\odot$ 
and NFW concentration c=5. The halo virial speed is $v_{vir} \approx$ 477 km/s. 
}
\label{fig:den_peter_cluster}
\end{figure}

For relevant lifetimes ($\Gamma^{-1} \gtrsim H_0^{-1}$), dark matter halos begin with the same density profiles 
as in the standard CDM model. Their density distributions can be well described by \citet{navarro_etal97} (NFW) profiles.  
As the DDM decays, the kinetic energy of dark matter particles
will change because SDM particles receive a small kick velocity 
from their parent particles.  Assuming that we only consider decay processes 
with $f \ll 1$, the mass of the parent and daughter particles will be nearly identical.  
As discussed in \citet{Sanchez-Salcedo_03}, {\em on average} the net effect of decays is to impart an amount of 
energy $\Delta E \approx m v^2_k/2$ on the dark matter, independent of the initial velocity. 
The changes in average kinetic energy will result in 
changes in particle orbits, causing an expansion of dark matter halos and a shallowing of dark 
matter profiles.

To demonstrate the effect of density profile modification, we adopt a two-step 
calculation.  Assume that DDM particles in halos follow circular orbits prior to any significant DDM decays.  
The particles orbit in the gravitational potential of the NFW halo, 
which can be approximately described by a power law $v_c(r)=v_0 (r/r_0)^{1/2\beta}$ over any sufficiently 
small range of $r$. In a given time interval, a small fraction of DDM particles decay and their daughter SDM 
particles gain a small amount of energy $\Delta E \approx m v^2_k/2$.  In general, the 
daughter particles will move from circular orbits to elongated orbits, characterized by 
the new energy relative to the halo potential and an apocentric radius $r$.  Orbits 
in the NFW potential are not closed, rendering it a numerical problem to compute 
the time-averaged value of the radial coordinate of the daughter particle.  To 
obtain a simplistic estimate of the new radii the particles move to, we assume 
that the new average position of the daughter is similar to the radius of a circular 
orbit at the new value of the orbital energy.  This is {\em conservative} in the present context, 
because circular orbits in equilibrium are {\em least} susceptible to such expansion \cite{Sanchez-Salcedo_03}. 
In this assumption, the radial position of the daughter particles, $r'$, will be
\begin{equation}
\label{eq30} 
r' = \left( r^{1/\beta} + \frac{1}{2\beta +1} \left( \frac{v_{k} (R)}{v_0} \right)^2 r_0^{1/\beta}\right)^\beta.
\end{equation}

The model we have described is not self-consistent, so it is important to validate the 
basic predictions of the model against more complete calculations.  
To check the validity of this model, we compare our analytical calculation 
results with the N-body simulation results from \cite{peter_etal10}.
In Figure~\ref{fig:den_peter}, we plot density profiles for a dark matter halo with mass 
$M_{\rm h} = 10^{12}\, \mathrm{M}_\odot$ and an initial NFW concentration parameter $c=5$ for several 
different lifetimes and kick velocities.  \citet{peter_etal10} computed the profiles of 
dark matter halos in the same model using N-body simulations that accounted for 
the dark matter decays.  Fig.~\ref{fig:den_peter} above is the same as Figure~1 
in \citet{peter_etal10} save for the fact that we have computed modified halo 
profiles according to the analytic model described in this section.  
A comparison of the two figures reveals that the analytic model and 
the numerical simulations are in remarkable agreement for all models 
with $v_k \lesssim 200$~km/s and $\Gamma^{-1} \gtrsim 10$~Gyr.  
There are several possible explanations for the inconsistencies that 
arise when $v_k > 200$~km/s and $\Gamma^{-1}< 10$~Gyr.  
One is that when changes to the gravitational potential are not small,
the final gravitational potential is sufficiently different from the 
initial gravitational potential that the initial potential cannot be used 
to approximate the new positions of the SDM particles. 
Another possibility is that typical circular orbits no 
longer provide useful approximations for the degree of halo expansion. 
As discussed in \cite{peter_etal10}, where they look at velocity anisotropy of their simulated halos, 
they found that the orbits become radially biased at the halo outskirts.  Moreover, $v_k = 200$~km/s 
is considerable compared to the virial velocity of a $M_{\rm h}=10^{12}\, \mathrm{M}_{\odot}$ halo, 
so it is not surprising that those halos are not in dynamical equilibrium 
for large $v_k$ and small lifetime.  These simulation results show that the assumptions 
of our simple model are violated in the regime of high kick velocity and low lifetime.  
As we show in \S~\ref{section:results}, our primary results in which the nonlinear 
model is used correspond to $v_k < 200$~km/s and lifetimes $\Gamma^{-1} > 100$~Gyr, so our use of 
this model for a first foray into this regime is justified.

Unlike \citet{peter_etal10}, we are interested in cosmological weak lensing as our observable.  The halo mass 
most relevant to weak lensing lie in the range $M_{\rm h} \approx 10^{13}-10^{14} M_{\odot}$ \cite{zentner_etal08}.  
Such halos have significantly larger virial velocities than the $10^{12}\, \mathrm{M}_{\odot}$ halos 
considered above.  Typical virial velocities of these larger halos lie in the range 
$v_{vir} \approx 280 - 600$~km/s.  This suggests that our model can be used at larger 
$v_k$ than the value $v_k \approx 200$~km/s that we arrived at by comparing to simulations 
of a $10^{12}\, \mathrm{M}_{\odot}$ halo above, because these kicks represent a smaller fraction 
of the potential well depth. For instance, \citet{peter_etal10} pointed out that the 
cluster mass function is insensitive to $v_k \lesssim$ 500 km/s, because the typical virial speeds 
clusters are $v_k \gtrsim$ 600 km/s.  For completeness, we show the corresponding density profile modifications for 
these group- and cluster-sized halos in Figure~\ref{fig:den_peter_cluster}.  We will show in \S~\ref{section:results} 
that our calculations are only sensitive to DDM parameters that result in density profiles with mild changes.

We include this effect in our nonlinear halo model calculation by giving all 
recomputing halo profiles as described above prior to computing lensing 
correlations.  We modified halo profiles by assuming initial halos 
with the same profiles, including concentrations, as their concordance 
$\Lambda$CDM counterparts and implementing the above model on these halos.
The remainder of our halo model implementation follows the approach 
we used in Ref.~\cite{wang_zentner10}, so we do not repeat it.  
Ideally, one would treat nonlinear corrections to structure growth 
using program of cosmological numerical simulations.  However, 
we place such a study outside the scope of the present work as 
our initial aim is to estimate the constraining power of forthcoming 
surveys.  In this manner, we estimate the fruit that a computationally-intensive 
numerical simulation program may bear on the problem of unstable dark matter.

\section{Forecasting Methods}
\label{section:fisher}

The Fisher Information Matrix provides a simple estimate of the parameter 
covariance given data of specified quality.  The Fisher matrix has 
been utilized in numerous, similar contexts in the cosmology literature 
\citep{jungman_etal96,tegmark_etal97,seljak97,hu99,kosowsky_etal02,huterer_takada05,
zentner_etal08,kitching_etal08,peter09,bernstein_huterer10,wang_zentner10}, 
so we give only a brief review of important results and 
the caveats in our particular application.  We have confirmed the 
validity of the Fisher matrix approximation in models of unstable 
dark matter using Monte Carlo methods as described in Ref.~\cite{wang_zentner10}.

The Fisher matrix of observables in Eq.~(\ref{eq:pkij}), subject to covariance 
as in Eq.~(\ref{cab}), can be written as
\beq
\label{eq:Fij}
F_{{i}{j}}=\sum_{\ell=\ell_{min}}^{\ell_{max}}(2\ell+1)f_{sky}\sum_{A,B} 
\frac{\partial P_{\kappa,A}}{\partial p_{i}}[C^{-1}]_{AB} 
\frac{\partial P_{\kappa,B}}{\partial p_{j}}+F_{ij}^P
\eeq
where the indices A and B run over all $\ntomo (\ntomo + 1)/2$ spectra and 
cross spectra, the $p_{i}$ are the parameters 
of the model, $f_{sky}$ is the fraction of the sky imaged 
by the experiment, and $\ell_{min}=2f_{sky}^{-1/2}$ is the smallest 
multipole constrained by the experiment.  
$F^P_{ij}$ is a prior Fisher matrix 
incorporating previous knowledge of viable regions of parameter space.  
We set $\ell_{max}=300$ for linear forecasts and $\ell_{max}=3000$ in our most ambitious nonlinear forecasts.  
On smaller scales (higher $\ell$), various assumptions, such as 
the Gaussianity of the lensing field, break down 
\cite{white_hu00,cooray_hu01,vale_white03,dodelson_etal06,semboloni_etal06,zentner_etal08,rudd_etal08}.  
To be conservative, we explore modest priors on 
each parameter independently, so that 
$F_{ij}^P=\delta_{ij}/(\sigma^P_i)^2$, where 
$\sigma^P_i$ is the 1$\sigma$ prior on parameter $p_{i}$.  
The forecast, 1$\sigma$, marginalized constraint on parameter 
$p_{i}$ is $\sigma(p_{i})=\sqrt{[F^{-1}]_{ii}}$.

Our DDM model has two independent parameters, namely decay rate 
$\Gamma$ (or lifetime, $\Gamma^{-1}$) and mass loss fraction $f$ 
(which is related to $v_k$ via $v_k = fc$).  
Either one of these parameters can independently be tuned 
to render the effects of DDM negligible, so it is not 
useful to marginalize over one parameter to derive constraints 
on the other.  In what follows, we choose to illustrate the 
effectiveness of lensing to constrain DDM by fixing lifetime and 
quoting possible constraints on $f$.  
Other than the mass loss fraction $f$, 
we also consider six cosmological parameters that we expect to 
modify weak lensing power spectra at significant levels and 
to exhibit partial degeneracy with our model parameters.  
We construct our forecasts for DDM lifetime bounds after 
marginalizing over the remaining parameters.  Our six additional 
parameters and their fiducial values (in parentheses) are the 
dark energy density $\Omega_{\Lambda}\ (0.74)$, the present-day dark 
matter density, $\omega_{\rm DM}=\Omega_{\rm DM} h^2\ (0.11)$, 
the baryon density $\omega_b=\Omega_b h^2\ (0.023)$, 
tilt parameter $n_s (0.963)$, the natural logarithm of the 
primordial curvature perturbation normalization $\ln(\Delta^2_R)\ (-19.94)$, 
and the sum of the neutrino masses $\sum_{i} m_{\nu_i}\ (0.05\ \mathrm{eV})$.  
This choice of fiducial model implies a small-scale, low-redshift power spectrum 
normalization of $\sigma_8 \simeq 0.82$.  The optical depth to reionization 
has a negligible effect on the lensing spectra on scales of 
interest, so we do not vary it in our analysis.

We take priors on our cosmological parameters of $\sigma(\omega_m)$ = 0.007, 
$\sigma(\omega_b)$ = $1.2 \times 10^{-3}$, $\sigma(\ln \Delta_R^2)$ = 0.1,
$\sigma(n_s)$ = 0.015, and $\sigma(\Omega_{\Lambda})$ = 0.03. 
We assume no priors on DDM model parameters or 
neutrino mass.  Our fiducial model is motivated by the WMAP seven-year result  
and our priors represent marginalized uncertainties on these parameters based on 
the WMAP seven-year data \citep{Komatsu_etal11}.  These priors are very {\em conservative} 
and allow for weaker constraints on DDM than would be expected from future data, 
where stronger priors may be available.  To estimate the potential power 
of lensing constraints on DDM when stronger cosmological constraints are 
available, we also explore prior constraints on these parameters 
at the level expected from the Planck mission\footnote{{\tt http://www.esa.int/planck}} 
using the entire Planck prior Fisher matrix of Ref.~\citep{hu_etal06}.  
Of course, using published priors from other analyses is not self-consistent 
because these priors were derived in analyses that assume stable dark 
matter, but for relevant lifetimes the dark matter decays should cause only 
subtle alterations to the cosmic microwave background anisotropy spectrum 
so this analysis should approximate a self-consistent simultaneous 
analysis of all data.

In some cases, we will estimate {\em nonlinear} power spectra 
in models with significant neutrino masses.  In such cases, 
we follow the empirical prescription established in previous studies 
(e.g., Refs.~\cite{Hannestad_etal06,kitching_etal08,Ichiki_etal09}) 
and take 
\beq
\label{mnmp}
P_{m}(k)=\left[f_{\nu}\sqrt{P^{\rm lin}_{\nu}(k)}+f_{b+DM}\sqrt{P^{NL}_{b+DM}(k)}\right]^2
\eeq
where 
\begin{subequations}
\label{mne}
\begin{align}
f_{\nu} &= { \Omega_{\nu}\over \Omega_{m}}\mathrm{,}\\
f_{\rm b+DM} &= { \Omega_{\rm DM} + \Omega_{b} \over \Omega_{m}}\mathrm{,}
\end{align}
\end{subequations}
$P^{\rm lin}_{\nu}(k)$ is the linear power spectrum of neutrinos, 
and $P^{\rm NL}_{\rm b+DM}(k)$ is the nonlinear power spectrum 
evaluated for baryons and dark matter only. 
However, we note that 
recent work has questioned the robustness of this treatment of neutrino 
mass using direct numerical simulations \cite{Bird_etal11} and 
perturbation theory \cite{Saito_etal08}, so it may become 
necessary to revisit this aspect of the modeling of power spectra prior to the 
availability of observational data.

We explore possible constraints from a variety of forthcoming data sets.  
We consider the Dark Energy Survey (DES) 
as a near-term imaging survey that could provide requisite data for this test.  
We model DES by taking a fractional sky coverage of $f_{sky}=0.12$ and 
with $\bar{n}=15/\mathrm{arcmin}^2$.  
Second, we consider a class of future ``Wide'' surveys as may be 
carried out by the Large Synoptic Survey Telescope 
(LSST)\citep{lsstbook} or 
Euclid \citep{eicbook}.  
We model these Wide surveys with $f_{sky}=0.5$ and 
$\bar{n}=50/\mathrm{arcmin}^2$.  
Lastly, we consider a comparably 
narrow, deep imaging survey.   
We refer to such a survey as a ``Deep'' survey and 
model it with $f_{sky}=0.05$ and $\bar{n}=100/\mathrm{arcmin}^2$. 
Such a survey may be more typical of a space-based mission 
similar to the proposed Wide-Field InfraRed Survey Telescope (WFIRST).  
In all cases, we take 
$\sqrt{\langle \gamma^2 \rangle}=0.2$ and assume particular 
shape measurement errors from each experiment are 
encapsulated in their effective number densities, in accord with 
recent conventional practice in this regard.  Our results are 
relatively insensitive to number density because shot noise 
does not dominate cosmic variance on the scales we consider, and 
our linear constraints are completely insensitive to the choice 
of galaxy number density over a wide range.

\begin{figure}[ht]
\includegraphics[height=7.5cm]{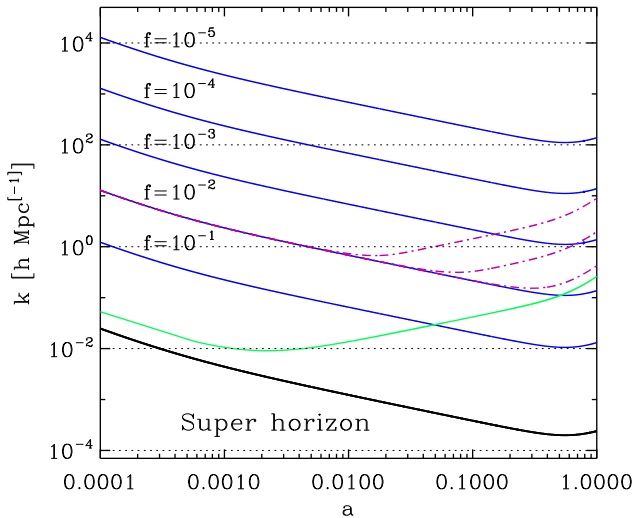}
\caption{ 
Free-streaming scale as a function of scale factor. The blue lines show free-streaming scales 
for lifetime much greater than the age of universe ( $>$ 100 Gyr) for several different
mass loss fractions. The dash-dotted magenta lines are for $f=10^{-2}$ and three lifetimes.  
From top-to-bottom at right, these are 0.01 Gyr, 0.1 Gyr, and 1 Gyr.  
The green line is the free-streaming scale for massive neutrino with $m_{\nu,i}$ =0.4 eV. 
Structure grows on scales between the free-streaming scale and
horizon.  On scales smaller than free-streaming scale ($k > k_{FS}$), 
structure growth is suppressed. 
}
\label{fig:ks}
\end{figure}

\begin{figure*}[ht]
\includegraphics[height=7.0cm]{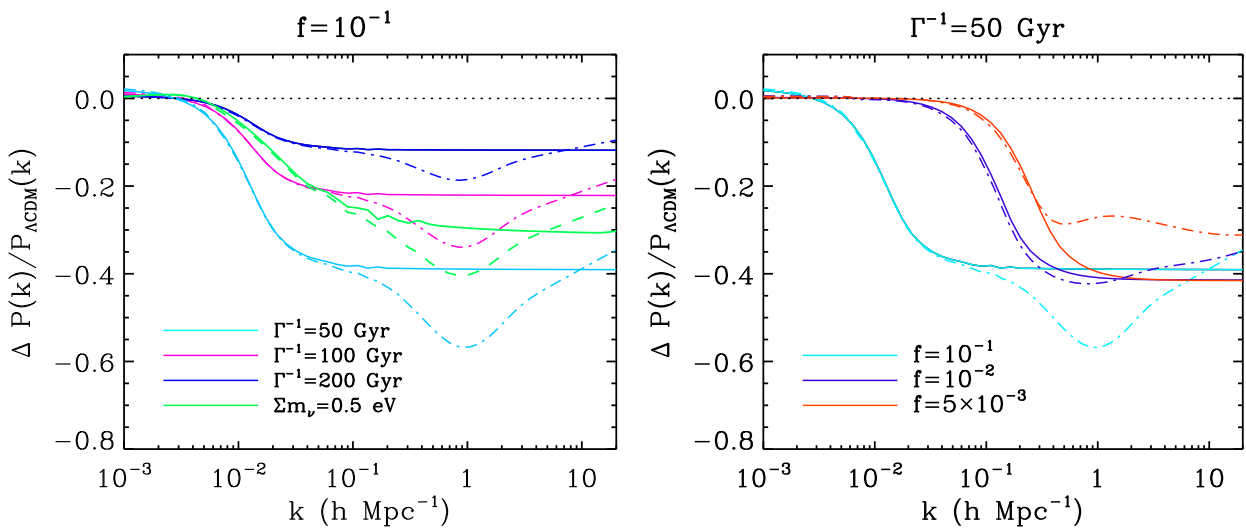}
\caption{ 
Fractional difference between matter power spectrum for standard $\Lambda$CDM 
and a decaying dark matter model evaluated at $z=0$. \textit{Left}: The effect of varying the 
DDM lifetime at fixed mass-loss fraction, $f=10^{-1}$.  Solid curves show the 
linear theory predictions, and dash or dash-dot lines show predictions that include the 
nonlinear corrections implemented via the halo model. The green lines show the spectrum in a 
$\Lambda$CDM with massive neutrinos, $\Sigma m_{\nu}=$0.5 eV, for comparison. 
\textit{Right}: The effect of varying mass-loss fraction $f$, at a fixed 
lifetime of $\Gamma^{-1}=50$~Gyr.
}
\label{fig:mps_all}
\end{figure*}

\section{Results}
\label{section:results}

There are several effects of DDM on lensing power spectra at low redshift.  
First, decays change the cosmological energy density. This change alters 
both structure growth and distance. Further, decaying dark matter results in 
significant free-streaming of daughter SDM particles.  While each of these effects 
can be important, for models near the limit of what may be constrained by 
lensing surveys, it is the effect of free-streaming that largely determines 
the lensing power spectra. In this case, the free-streaming velocity of SDM suppresses 
structure growth on scales smaller than free-streaming scale, an effect similar to that 
caused by massive neutrinos. In \S~\ref{section:fss} we show the behavior of 
the free-streaming scale in our model and compare it with massive neutrinos.  
We follow this by showing the alterations to lensing power spectra in 
\S~\ref{section:PS}.  Our final results are the forecast constraints on 
DDM, which we give in \S~\ref{section:forecast}.

\subsection{Free-streaming Scale}
\label{section:fss}

In the standard cosmological scenario, matter density fluctuations at a particular scale 
grow once the scale enters the horizon ($k>H$) during the matter-dominated epoch. 
However, species with non-negligible primordial velocities will be able to escape 
the potential wells and suppress the formation of structure. The scale that corresponds 
to this effect is the free-streaming scale $k_{FS}$, which can be defined as 
\begin{equation}
\label{kfs} 
k_{FS}(a)=\sqrt{3 \over 2} { \mathcal{H}(a) \over c_s(a)},
\end{equation}
where $\mathcal{H}(a)=a{\dd a/ \dd \tau}$ and $\mathcal{H}^{-1}$ is the 
comoving horizon scale.

We show the evolution of free-streaming scale of SDM particles as a function of 
scale factor in Figure~\ref{fig:ks} for several mass loss fractions $f$ and lifetimes.  
As discussed in \cite{Aoyama_etal11}, the behavior of the free-streaming scale of DDM
can be divided into two regimes. When the decay process is still occurring, corresponding to 
cosmological times less than the decay lifetime, daughter particles with the same
physical momentum are continuously created so that the sound speed stays approximately 
the same. In this case, the evolution of free-streaming scale will simply trace 
the evolution of horizon. If decays have ceased, which will happen when $\Gamma^{-1} < H_0^{-1}$, 
the sound speed will decrease as $c_s \propto a^{-1}$.  The free-streaming scale shrinks as the 
initial velocities are redshifted away.  This effect also happens to massive neutrinos as they become 
non-relativistic. At early times the neutrino free-streaming scale traces the horizon so long as the 
neutrinos have relativistic velocities. In Figure~\ref{fig:ks} we
can see that after neutrinos become non-relativistic, at 
$a_{nr} \simeq 1.3 \times 10^{-3}\, (0.4\,\mathrm{eV}/m_{\nu})$, 
their free-streaming scale varies as $ k_{FS} \propto a^{1/2}$ during 
matter domination, 
which is identical to free-streaming in the small lifetime limit of DDM.

\subsection{Weak lensing Power Spectrum}
\label{section:PS}

As we mentioned above, DDM affects lensing power spectra in two respects.  First, 
the power spectra for potential and density fluctuations are modified by the 
free streaming of the daughter SDM particles.  At $k \gtrsim k_{FS}$, structure 
growth is suppressed. Second, the matter density is reduced as decays occur, 
slightly suppressing the late-time growth of structure.  
In the left panel in Figure~\ref{fig:mps_all}, we show that significant 
decrements in power occur at roughly the same 
scale, $k \gtrsim10^{-2} h\,\mathrm{Mpc}^{-1}$ for a variety of 
lifetimes, so long as the lifetime $\Gamma^{-1} \gg H_0^{-1}$ 
(the regime most relevant to our work).  
This suppression is due to free streaming and indeed, 
the scale on which the suppression occurs agrees with 
the estimates of the free-streaming scale shown in 
Fig.~\ref{fig:ks}.  The right panel of Fig.~\ref{fig:mps_all} 
illustrates that the scale of suppression is determined by the 
mass-loss fraction $f$, in the limit that $\Gamma^{-1} \gg H_{0}^{-1}$.  
In models with larger $f$, the velocities of the daughter 
SDM particles are higher, so at fixed lifetime, they free-stream 
greater distances.  Both panels in Fig.~\ref{fig:mps_all}  
show a small increment in power on large scales for models 
with small lifetimes ($\Gamma^{-1} \lesssim 50$~Gyr) 
and larger mass-loss fractions ($f \gtrsim 0.1$).  
This delineates the parameter regime for which the 
overall change in the energy budget begins to have a 
non-negligible effect on fluctuation growth.  The small 
increment on large scales in these cases enforces a fixed 
observed CMB normalization.

Notice in the left panel of Fig.~\ref{fig:mps_all} that with $f \sim 10^{-1}$, the free-streaming suppression 
is similar to that induced by massive neutrinos with the sum of the neutrino masses 
$\Sigma m_{\nu} \approx 0.5$~eV.  This suggests that neutrinos may be degenerate with 
DDM, and this would be the case if it were not possible to probe a wide range of 
length scales and redshifts.  In practice, we find that massive neutrinos are 
distinguishable from DDM for two reasons.  First, the differences in scale dependence exhibited 
in Fig.~\ref{fig:mps_all} give a possible handle with which to separate the two.  More importantly, 
the redshift dependence of the power spectrum differs in the two models.  This is most easily seen 
in Fig.~\ref{fig:ks}.  The evolution of the free-streaming scale of massive neutrinos and the 
free-streaming scale of DDM differs significantly.  Deep, large-scale survey data that enable 
probes of structure at a variety of redshifts between $0 \lesssim z \lesssim 3$, as is expected 
of forthcoming surveys, break the potential degeneracy between massive neutrinos and DDM.

\begin{figure}[t]
\includegraphics[height=7.2cm]{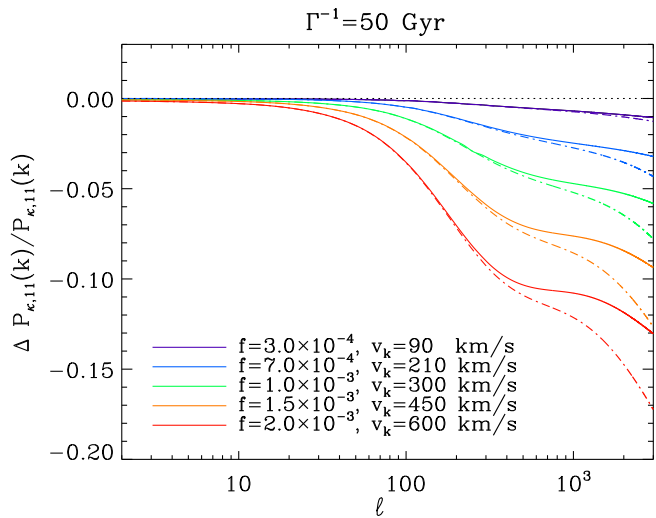}
\caption{ 
Fractional difference of auto convergence lensing power spectrum between 
standard $\Lambda$CDM model and decaying dark matter model from first tomographic redshift bin 
(lensing source galaxies between $0 < z_p < 0.6$, where $z_p$ is photometric, and not necessarily 
true redshift).  Solid lines are calculated using halo model with NFW profiles. These lines include 
the alteration of the linear power spectrum on large-scales and the reduction in the abundance of 
dark matter halos due to free-streaming.  However, halos are assumed to have the same profiles as 
they would in standard $\Lambda$CDM.  The Dash-dotted lines include the nonlinear corrections to 
halo density profiles.
}
\label{fig:pkappa_m}
\end{figure}

The observed strength of gravitational lensing also has a dependence upon geometry, 
so differences in angular diameter distance may lead to modified lensing power 
spectra. Changes in relative partitioning of energy among relativistic and non-relativistic 
species will change the evolution of the angular diameter distance. However, we 
consider small mass loss fractions ($f\ll$1) and large lifetimes ($\Gamma^{-1} >> H_0^{-1}$), 
so angular diameter distances are altered only by negligible amounts and, although we 
account for these changes, they do not provide leverage on constraining DDM.  
For example, for a lifetime of $\Gamma^{-1}=50$~Gyr and $f=10^{-1}$, Figure~\ref{fig:mps_all} shows that 
the changes in matter power spectrum are approximately $40-60\%$ for 
$k \gtrsim 2\times10^{-2}\ h\,\mathrm{Mpc}^{-1})$.  
The corresponding changes in angular diameter distances are 
less than $0.1\%$.  The constraining power of weak lensing comes primarily from the 
suppression of structure growth.  Incidentally, this is a promising feature because the 
primary information used to constrain dark energy using lensing surveys is carried by 
the geometric piece of the lensing signal \cite{zhan_knox06,hearin_zentner09}.  

\citet{Aoyama_etal11} have considered constraints on unstable dark matter stemming from 
contemporary data on the cosmological distance-redshift relations.  Using constraints 
on the Hubble parameter, the baryon acoustic oscillation scale, and the angular positions 
of the cosmic microwave background anisotropy spectrum peaks, they have placed competitive 
constraints on such models.  \citet{Aoyama_etal11} find that $\Gamma^{-1} \gtrsim 0.1$~Gyr 
with $f \lesssim 3.5 \times 10^{-2}$ and that $\Gamma^{-1} \gtrsim 30$~Gyr when 
$f \sim 1$.

The Dash-dotted lines in Figure~\ref{fig:pkappa_m} exemplify the alterations to the small-scale 
lensing convergence power spectra incurred when we account for the altered halo profiles that 
result from dark matter decays.  As $f$ increases, kick velocities increase, and 
the fractional power decrement increases, as we should expect.  This additional suppression 
is confined to relatively small scales (large multipoles, $\ell \gtrsim 300$) for most of 
the parameter space of interest ($v_k \lesssim 200\ \mathrm{km/s}$ for $\Gamma^{-1} \lesssim 100\ \mathrm{Gyr}$).   

\begin{figure}[ht!]
\includegraphics[height=7.0cm]{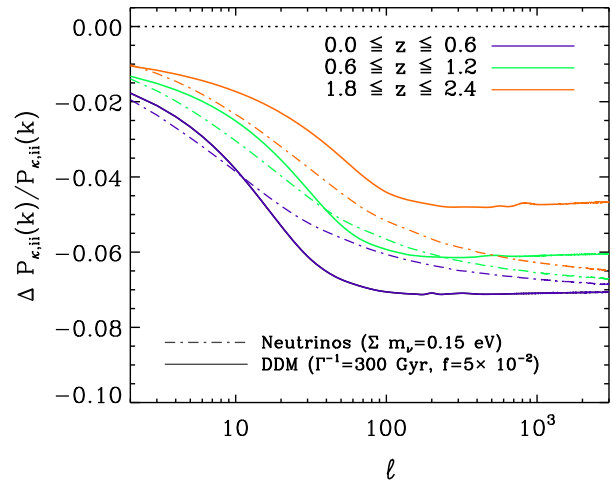}
\caption{ 
Comparison of the redshift evolution of decaying dark matter and massive neutrino lensing 
power spectra.  We plot fractional difference of auto convergence lensing power spectra between 
standard $\Lambda$CDM model and decaying dark matter (or massive neutrino) models in three tomographic 
redshift bins (labeled at the top).  For simplicity, we show only the linear power spectra in this plot, though 
spectra computed with our nonlinear model lead to a similar conclusion.
}
\label{fig:pkappa_m_z}
\end{figure}

As we pointed out in right panel in Figure~\ref{fig:mps_all}, DDM may partially mimic 
massive neutrinos if redshift evolution information in not accessible. 
In Figure~\ref{fig:pkappa_m_z}, we show a comparison of the redshift evolution of 
DDM and massive neutrino lensing power spectra in three tomographic redshift bins. 
Other than the difference in shapes, it is also evident that the DDM power spectra 
evolve significantly more than the spectra in massive neutrino models.  
The reason is that the decay process continuously deposits kinetic energy 
into the daughter dark matter distribution, in contrast to the neutrinos 
which have purely redshifting kinetic energy distributions.

\subsection{Forecast Constraints on DDM Model Parameters}
\label{section:forecast}

%
\begin{figure*}[ht]
\includegraphics[height=10cm]{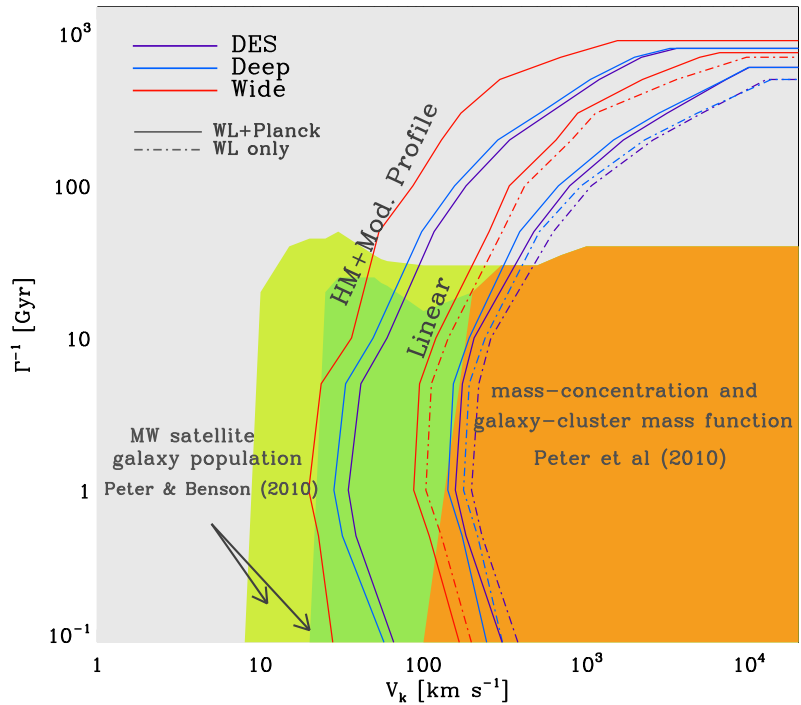}
\caption{
Comparison of DDM parameter exclusion contours from 
\citet{peter_etal10} (orange) and \citet{peter_benson10} (dark and light green) to 
those that may be achieved from weak gravitational lensing.  
The exclusions in \citep{peter_benson10} rely on a variety of assumptions 
regarding the merger history and formation histories of galaxies.  Roughly speaking 
these parameter regimes also correspond to parameter values for which the interpretation 
of the missing satellites problem may be significantly altered by unstable dark matter 
(see Ref.~\cite{peter_benson10} for details).  
The red, blue, and purple lines are the 1 $\sigma$ exclusion contours from our 
weak lensing forecasts assuming
"wide," "deep," and DES like galaxy imaging surveys respectively. 
The solid lines combine weak lensing with projected Planck constraints,
and the dot-dash lines are from weak lensing alone.  The less restrictive 
set of contour lines correspond to weak lensing constraints using scales 
on which linear theory is applicable ($\ell < 300$).  The more restrictive 
set of contours incorporate multipoles up to $\ell_{\rm max} = 3000$ and the 
theoretical calculation includes our nonlinear corrections to halo density profiles.
}
\label{fig:constraints}
\end{figure*}

To estimate of the power of weak lensing to constrain 
DDM, we adopt a variety of possible strategies.  
First, we consider constraints from data on scales where linear evolution 
of density fluctuations should be valid.  The value of this approach 
is that exploiting linear scales to constrain DDM does not require a 
simulation program to confirm or refine nonlinear models of structure 
formation in these models.  This can be done with contemporary theoretical 
knowledge of the phenomenology of these models.  Moreover, relatively 
large-scale constraints are less observationally challenging because they 
exploit data on scales where cosmic variance, rather than galaxy 
shape measurements, are the dominant error \cite{wang_zentner10}.  
In both cases, these constraints are conservative so we should expect 
that forthcoming lensing surveys designed to address dark energy 
should do {\em at least as well as our linear forecasts}.  To limit ourselves 
to linear scales, we take data on multipoles $\ell < 300$.  All of the constraints 
that we show in this section have been marginalized over the remaining cosmological 
parameters, including neutrino mass.

To show the maximum potential of lensing surveys, we consider measurements that 
extend into the mildly nonlinear regime, as is commonly done for dark energy 
forecasts.  The primary value of this extension is not that particular features 
in the power spectra induced by DDM are added to the data set.  Rather the primary 
improvement in constraints comes from an increase in the signal-to-noise with which 
the power suppression can be detected \cite{wang_zentner10}.  In this case, 
we include information on multipoles up to our quoted maximum multipole $\ell_{\rm max}=3000$ 
(see \S~\ref{section:fisher}).  Constraints on these scales will rely on reliable modeling of 
clustering on mildly nonlinear scales, so a comprehensive simulation program will be necessary 
to ensure the robustness of such constraints.  A comprehensive program is computationally-intensive 
and beyond the scope of our present paper, as part of our goal is to emphasize that such a 
large-scale numerical program may be interesting and useful.  

In Figure~\ref{fig:constraints} we display our forecast $1\sigma$ exclusion 
contours alongside a variety of other contemporary constraints.  
The most relevant contemporary constraints come from 
modifications to the structures of dark matter halos with 
virial velocities similar to the SDM kick velocities \cite{peter_etal10} (orange region).  
Additional constraints may be placed on unstable dark matter by examining 
the properties of the satellite galaxies of the Milky Way \cite{peter_benson10} (green regions).   
However, these constraints rely on a variety of assumptions regarding the 
formation and evolution of relatively small galaxies.  Moreover, these 
constraints delineate a range of DDM parameters for which unstable 
dark matter may have a significant effect on the interpretation of the 
missing satellites problem.  As this type of DDM model has been invoked 
to mitigate the "missing satellite" problem, it should not be a 
surprise that stronger constraints also come from these types of observations.
As such, it is this parameter range for which it is 
most interesting to develop independent constraints on unstable dark matter and 
that is the purpose of our weak lensing study.  

As indicated in Figure~\ref{fig:constraints}, our most conservative, linear calculation 
can already give interesting constraints DDM that are competitive with contemporary 
bounds.  The largest advantage of lensing constraints will be that it can extend 
constraints on DDM lifetimes significantly, as is evident in Fig.~\ref{fig:constraints}.  
These forecasts are not dependent upon modeling nonlinear structure growth, 
so they constitute a robust lower limit to the constraining power of imaging surveys.  
Moreover, these constraints are not subject to any particular assumptions 
regarding the evolution and formation of galaxies, particularly the Milky Way 
satellite galaxies that are the subject of so much contemporary research.  
Comparing the linear constraints from the three types of surveys, 
the ``Deep'' survey provides slightly more restrictive constraints 
than DES. A ``Wide'' survey similar to LSST or 
Euclid has the potential to improve the constraints relative to DES by 
$\sim$40-60$\%$. 

The slope of the constraint contours turns over near lifetimes of a few Gyr.  
This turn over reflects the turn over in free streaming scale exhibited by 
the dash-dotted lines in Fig.~\ref{fig:ks}.  In models with 
$\Gamma^{-1} \ll H_0^{-1}$, the free-streaming scale at low redshift 
decreases with time.  Notice that including Planck priors yields only a 
marginal improvement on the forecast constraints, $\sim 15-40\%$ over the 
parameter ranges of interest.  Our nonlinear forecasts exhibit a similar 
sensitivity to Planck priors, though they are not depicted in Fig.~\ref{fig:constraints} 
in the interest of clarity.

Our more ambitious approach is to estimate constraints including data on 
mildly nonlinear scales.  As we have already mentioned, forecast constraints 
in this case are less robust because they rely on data in a regime on which 
systematic errors and galaxy shape noise, as opposed to cosmic variance, 
will likely be the largest errors and because theoretical models of 
structure growth on these scales will need to be validated using a 
large suite of cosmological structure formation simulations.  The 
most ambitious constraint forecasts in Fig.~\ref{fig:constraints} 
result from including multipoles $\ell \le \ell_{\rm max}=3000$ 
along with Planck priors on the standard cosmological parameters.  
These forecasts indicate that weak lensing constraints on $v_k$ that include 
mildly nonlinear scales may improve upon the minimal, linear constraints 
by nearly an order of magnitude.  In this case, it is clear that 
weak lensing may provide an independent and competitive constraint on 
DDM and that lensing constraints will extend to significantly longer 
lifetimes than contemporary bounds.

\section{Discussion and Conclusion}
\label{section:conclusion}

In this paper we have explored the effect of decaying dark matter on the 
large-scale matter distribution and the power of future weak lensing surveys 
to place constraints on DDM lifetimes and mass splittings.  The mass difference 
is parameterized by the velocity kick ($v_k$) that the daughter, stable dark matter particles 
receive upon the decay of the heavier, parent DDM.  DDM leads to a suppression 
of matter clustering on scales below the free-streaming scale of the stable, 
daughter dark matter particles and this suppression can be probed with 
data from galaxy imaging surveys.  

Our most conservative constraint forecasts result from considering lensing over large scales on which 
linear theory should be valid.  In this case, best limits which come from a ``Wide'' survey, 
similar to Euclid or LSST.  These surveys may exclude $v_k \gtrsim 90$~km/s for 
$\Gamma^{-1} \sim  1-5$~Gyr, a result that is competitive with contemporary 
constraints \cite{peter_etal10}.  Lensing improves upon contemporary constraints most markedly 
for large decay lifetimes ($\Gamma^{-1} > H_0^{-1}$).  In this regime, 
lensing constraints are significantly more restrictive than contemporary 
bounds from halo structure.  In the relatively near-term, the 
DES will be able to place limits of $v_k \gtrsim 160$~km/s for 
$\Gamma^{-1}\sim 1-5$~Gyr. Achieving constraints at this level should 
be achievable.  First, the lensing surveys we study are under development 
to study dark energy already.  Moreover, these constraints assume that 
we restrict attention only to relatively large scales on which linear 
perturbation theory can be used to predict lensing power, so no 
additional theoretical effort will be necessary.

It may be possible to derive more restrictive lensing constraints on unstable 
dark matter by considering the mildly nonlinear scales that are commonly considered 
as part of the program to constrain dark energy.  Including multipoles up to 
$ell \sim 10^3$ increases constraining power by boosting the signal-to-noise with 
of the weak lensing signal on scales that are sensitive to the dynamics of the 
dark matter.  Exploiting such scales will 
rely on an exhaustive simulation program to understand nonlinear clustering in 
DDM models, similar to the simulation program that is being performed in support 
of dark energy probes \cite{heitmann_etal08}, so significant additional theoretical 
work will be necessary.  Nevertheless, the payoff may be significant.  In order to 
estimate the ambitious constraints that may be achieved from such a data analysis, 
we have implemented nonlinear corrections to lensing power using the standard 
halo model coupled with a simple model for the modification of halo density structures 
due to decaying dark matter.  

In our most ambitious forecasts, we find that weak lensing may constrain the 
mass splitting of the DDM nearly an order of magnitude more restrictively than 
implied by our linear scale analysis.  These results suggest that weak lensing 
surveys will be sensitive to $v_k \sim 10$~km/s for lifetimes $\Gamma^{-1}\lesssim 10$~Gyr 
for all the survey types that we have considered.  These constraints are interesting 
because they restrict parameters for which considerable effects due to DDM may be 
evident in the Milky Way satellite galaxy population.  It may be possible to 
achieve similar constraints depending upon a variety of assumptions regarding the 
formation process of these satellite galaxies \cite{peter_benson10}, but lensing 
provides a complementary constraint using data on distinct length scales.  

We have demonstrated that measurements of the large-scale matter distribution through a 
weak lensing survey will be a powerful probe of decaying dark matter.  This probe is 
valuable for several reasons.  First, such surveys as PanSTARRS, LSST, DES, Euclid, and 
WFIRST are already being undertaken as part of the effort to constrain dark energy.  
The survey requirements specified by the dark energy program are the same that we 
assume here, so no additional observational work will be necessary.  Moreover, we 
have shown that such measurements can provide independent, competitive constraints 
on models of DDM that could alter our interpretation of the small-scale problems 
of the standard cosmological model, particularly the missing satellites problem.  
In fact, we have demonstrated that lensing will probe DDM models with lifetimes 
that exceed contemporary bounds by an order of magnitude.  Our most ambitious 
constraint forecasts rely upon the development of accurate and precise models 
of matter clustering in models of unstable dark matter.  This will likely 
require a significant simulation effort to ensure the robustness of any 
constraints derived from forthcoming data.  It is our hope that this
proof-of-concept work will motivate more detailed numerical studies of 
unstable dark matter models as well as additional possible constraints from 
related observations.

%
%
\vspace*{12pt}
\begin{acknowledgments}
We are grateful to Mickey Abbott, Andrew Hearin, Arthur Kosowsky, 
Jeff Newman, Annika Peter, and Chris Purcell for useful comments and 
discussions.  We thank Annika Peter for providing the constraint 
contours from her work depicted in Fig.~\ref{fig:constraints}.  
This work was funded by the Pittsburgh Particle Physics, Astrophysics, 
and Cosmology Center (PITT PACC) at the University of Pittsburgh, 
and the National Science Foundation through grant PHY 0968888.  
We thank Uros Seljak and Mathias Zaldarriage 
for use of the publicly-available {\tt CMBFAST} code.  
This work made use of the NASA Astrophysics Data System.

\end{acknowledgments}


\bibliography{wlddm}

\begin{thebibliography}{75}
\expandafter\ifx\csname natexlab\endcsname\relax\def\natexlab#1{#1}\fi
\expandafter\ifx\csname bibnamefont\endcsname\relax
  \def\bibnamefont#1{#1}\fi
\expandafter\ifx\csname bibfnamefont\endcsname\relax
  \def\bibfnamefont#1{#1}\fi
\expandafter\ifx\csname citenamefont\endcsname\relax
  \def\citenamefont#1{#1}\fi
\expandafter\ifx\csname url\endcsname\relax
  \def\url#1{\texttt{#1}}\fi
\expandafter\ifx\csname urlprefix\endcsname\relax\def\urlprefix{URL }\fi
\providecommand{\bibinfo}[2]{#2}
\providecommand{\eprint}[2][]{\url{#2}}

\bibitem[{\citenamefont{{Jungman}
  et~al.}(1996{\natexlab{a}})\citenamefont{{Jungman}, {Kamionkowski}, and
  {Griest}}}]{jungman_etal96b}
\bibinfo{author}{\bibfnamefont{G.}~\bibnamefont{{Jungman}}},
  \bibinfo{author}{\bibfnamefont{M.}~\bibnamefont{{Kamionkowski}}},
  \bibnamefont{and} \bibinfo{author}{\bibfnamefont{K.}~\bibnamefont{{Griest}}},
  \bibinfo{journal}{\physrep} \textbf{\bibinfo{volume}{267}},
  \bibinfo{pages}{195} (\bibinfo{year}{1996}{\natexlab{a}}),
  \eprint{arXiv:hep-ph/9506380}.

\bibitem[{\citenamefont{{Griest} and
  {Kamionkowski}}(2000)}]{griest_kamionkowski00}
\bibinfo{author}{\bibfnamefont{K.}~\bibnamefont{{Griest}}} \bibnamefont{and}
  \bibinfo{author}{\bibfnamefont{M.}~\bibnamefont{{Kamionkowski}}},
  \bibinfo{journal}{\physrep} \textbf{\bibinfo{volume}{333}},
  \bibinfo{pages}{167} (\bibinfo{year}{2000}).

\bibitem[{\citenamefont{{Bertone} et~al.}(2005)\citenamefont{{Bertone},
  {Hooper}, and {Silk}}}]{bertone_etal05}
\bibinfo{author}{\bibfnamefont{G.}~\bibnamefont{{Bertone}}},
  \bibinfo{author}{\bibfnamefont{D.}~\bibnamefont{{Hooper}}}, \bibnamefont{and}
  \bibinfo{author}{\bibfnamefont{J.}~\bibnamefont{{Silk}}},
  \bibinfo{journal}{\physrep} \textbf{\bibinfo{volume}{405}},
  \bibinfo{pages}{279} (\bibinfo{year}{2005}), \eprint{arXiv:hep-ph/0404175}.

\bibitem[{\citenamefont{{Komatsu} et~al.}(2011)\citenamefont{{Komatsu},
  {Smith}, {Dunkley}, {Bennett}, {Gold}, {Hinshaw}, {Jarosik}, {Larson},
  {Nolta}, {Page} et~al.}}]{Komatsu_etal11}
\bibinfo{author}{\bibfnamefont{E.}~\bibnamefont{{Komatsu}}},
  \bibinfo{author}{\bibfnamefont{K.~M.} \bibnamefont{{Smith}}},
  \bibinfo{author}{\bibfnamefont{J.}~\bibnamefont{{Dunkley}}},
  \bibinfo{author}{\bibfnamefont{C.~L.} \bibnamefont{{Bennett}}},
  \bibinfo{author}{\bibfnamefont{B.}~\bibnamefont{{Gold}}},
  \bibinfo{author}{\bibfnamefont{G.}~\bibnamefont{{Hinshaw}}},
  \bibinfo{author}{\bibfnamefont{N.}~\bibnamefont{{Jarosik}}},
  \bibinfo{author}{\bibfnamefont{D.}~\bibnamefont{{Larson}}},
  \bibinfo{author}{\bibfnamefont{M.~R.} \bibnamefont{{Nolta}}},
  \bibinfo{author}{\bibfnamefont{L.}~\bibnamefont{{Page}}},
  \bibnamefont{et~al.}, \bibinfo{journal}{\apjs}
  \textbf{\bibinfo{volume}{192}}, \bibinfo{pages}{18} (\bibinfo{year}{2011}),
  \eprint{arXiv:1001.4538}.

\bibitem[{\citenamefont{{Klypin} et~al.}(1999)\citenamefont{{Klypin},
  {Kravtsov}, {Valenzuela}, and {Prada}}}]{klypin_etal99b}
\bibinfo{author}{\bibfnamefont{A.}~\bibnamefont{{Klypin}}},
  \bibinfo{author}{\bibfnamefont{A.~V.} \bibnamefont{{Kravtsov}}},
  \bibinfo{author}{\bibfnamefont{O.}~\bibnamefont{{Valenzuela}}},
  \bibnamefont{and} \bibinfo{author}{\bibfnamefont{F.}~\bibnamefont{{Prada}}},
  \bibinfo{journal}{\apj} \textbf{\bibinfo{volume}{522}}, \bibinfo{pages}{82}
  (\bibinfo{year}{1999}), \eprint{astro-ph/9901240}.

\bibitem[{\citenamefont{{Moore} et~al.}(1999)\citenamefont{{Moore}, {Ghigna},
  {Governato}, {Lake}, {Quinn}, {Stadel}, and {Tozzi}}}]{moore_etal99}
\bibinfo{author}{\bibfnamefont{B.}~\bibnamefont{{Moore}}},
  \bibinfo{author}{\bibfnamefont{S.}~\bibnamefont{{Ghigna}}},
  \bibinfo{author}{\bibfnamefont{F.}~\bibnamefont{{Governato}}},
  \bibinfo{author}{\bibfnamefont{G.}~\bibnamefont{{Lake}}},
  \bibinfo{author}{\bibfnamefont{T.}~\bibnamefont{{Quinn}}},
  \bibinfo{author}{\bibfnamefont{J.}~\bibnamefont{{Stadel}}}, \bibnamefont{and}
  \bibinfo{author}{\bibfnamefont{P.}~\bibnamefont{{Tozzi}}},
  \bibinfo{journal}{\apjl} \textbf{\bibinfo{volume}{524}}, \bibinfo{pages}{L19}
  (\bibinfo{year}{1999}), \eprint{arXiv:astro-ph/9907411}.

\bibitem[{\citenamefont{{de Blok} and {Bosma}}(2002)}]{deBlok_etal02}
\bibinfo{author}{\bibfnamefont{W.~J.~G.} \bibnamefont{{de Blok}}}
  \bibnamefont{and} \bibinfo{author}{\bibfnamefont{A.}~\bibnamefont{{Bosma}}},
  \bibinfo{journal}{\aap} \textbf{\bibinfo{volume}{385}}, \bibinfo{pages}{816}
  (\bibinfo{year}{2002}), \eprint{arXiv:astro-ph/0201276}.

\bibitem[{\citenamefont{{Simon} et~al.}(2005)\citenamefont{{Simon}, {Bolatto},
  {Leroy}, {Blitz}, and {Gates}}}]{Simon_etal05}
\bibinfo{author}{\bibfnamefont{J.~D.} \bibnamefont{{Simon}}},
  \bibinfo{author}{\bibfnamefont{A.~D.} \bibnamefont{{Bolatto}}},
  \bibinfo{author}{\bibfnamefont{A.}~\bibnamefont{{Leroy}}},
  \bibinfo{author}{\bibfnamefont{L.}~\bibnamefont{{Blitz}}}, \bibnamefont{and}
  \bibinfo{author}{\bibfnamefont{E.~L.} \bibnamefont{{Gates}}},
  \bibinfo{journal}{\apj} \textbf{\bibinfo{volume}{621}}, \bibinfo{pages}{757}
  (\bibinfo{year}{2005}), \eprint{arXiv:astro-ph/0412035}.

\bibitem[{\citenamefont{{Kuzio de Naray} et~al.}(2008)\citenamefont{{Kuzio de
  Naray}, {McGaugh}, and {de Blok}}}]{KuziodeNaray_etal08}
\bibinfo{author}{\bibfnamefont{R.}~\bibnamefont{{Kuzio de Naray}}},
  \bibinfo{author}{\bibfnamefont{S.~S.} \bibnamefont{{McGaugh}}},
  \bibnamefont{and} \bibinfo{author}{\bibfnamefont{W.~J.~G.} \bibnamefont{{de
  Blok}}}, \bibinfo{journal}{\apj} \textbf{\bibinfo{volume}{676}},
  \bibinfo{pages}{920} (\bibinfo{year}{2008}), \eprint{0712.0860}.

\bibitem[{\citenamefont{{Col{\'{\i}}n}
  et~al.}(2000)\citenamefont{{Col{\'{\i}}n}, {Avila-Reese}, and
  {Valenzuela}}}]{Colin_etal00}
\bibinfo{author}{\bibfnamefont{P.}~\bibnamefont{{Col{\'{\i}}n}}},
  \bibinfo{author}{\bibfnamefont{V.}~\bibnamefont{{Avila-Reese}}},
  \bibnamefont{and}
  \bibinfo{author}{\bibfnamefont{O.}~\bibnamefont{{Valenzuela}}},
  \bibinfo{journal}{\apj} \textbf{\bibinfo{volume}{542}}, \bibinfo{pages}{622}
  (\bibinfo{year}{2000}), \eprint{arXiv:astro-ph/0004115}.

\bibitem[{\citenamefont{{Bode} et~al.}(2001)\citenamefont{{Bode}, {Ostriker},
  and {Turok}}}]{Bode_etal01}
\bibinfo{author}{\bibfnamefont{P.}~\bibnamefont{{Bode}}},
  \bibinfo{author}{\bibfnamefont{J.~P.} \bibnamefont{{Ostriker}}},
  \bibnamefont{and} \bibinfo{author}{\bibfnamefont{N.}~\bibnamefont{{Turok}}},
  \bibinfo{journal}{\apj} \textbf{\bibinfo{volume}{556}}, \bibinfo{pages}{93}
  (\bibinfo{year}{2001}), \eprint{arXiv:astro-ph/0010389}.

\bibitem[{\citenamefont{{Lovell} et~al.}(2011)\citenamefont{{Lovell}, {Eke},
  {Frenk}, {Gao}, {Jenkins}, {Theuns}, {Wang}, {Boyarsky}, and
  {Ruchayskiy}}}]{Lovell_etal11}
\bibinfo{author}{\bibfnamefont{M.}~\bibnamefont{{Lovell}}},
  \bibinfo{author}{\bibfnamefont{V.}~\bibnamefont{{Eke}}},
  \bibinfo{author}{\bibfnamefont{C.}~\bibnamefont{{Frenk}}},
  \bibinfo{author}{\bibfnamefont{L.}~\bibnamefont{{Gao}}},
  \bibinfo{author}{\bibfnamefont{A.}~\bibnamefont{{Jenkins}}},
  \bibinfo{author}{\bibfnamefont{T.}~\bibnamefont{{Theuns}}},
  \bibinfo{author}{\bibfnamefont{J.}~\bibnamefont{{Wang}}},
  \bibinfo{author}{\bibfnamefont{A.}~\bibnamefont{{Boyarsky}}},
  \bibnamefont{and}
  \bibinfo{author}{\bibfnamefont{O.}~\bibnamefont{{Ruchayskiy}}},
  \bibinfo{journal}{ArXiv e-prints}  (\bibinfo{year}{2011}),
  \eprint{arXiv:1104.2929}.

\bibitem[{\citenamefont{{Zentner} and {Bullock}}(2002)}]{zentner_bullock02}
\bibinfo{author}{\bibfnamefont{A.~R.} \bibnamefont{{Zentner}}}
  \bibnamefont{and} \bibinfo{author}{\bibfnamefont{J.~S.}
  \bibnamefont{{Bullock}}}, \bibinfo{journal}{\prd}
  \textbf{\bibinfo{volume}{66}}, \bibinfo{eid}{043003} (\bibinfo{year}{2002}),
  \eprint{arXiv:astro-ph/0205216}.

\bibitem[{\citenamefont{{Zentner} and {Bullock}}(2003)}]{zentner_bullock03}
\bibinfo{author}{\bibfnamefont{A.~R.} \bibnamefont{{Zentner}}}
  \bibnamefont{and} \bibinfo{author}{\bibfnamefont{J.~S.}
  \bibnamefont{{Bullock}}}, \bibinfo{journal}{\apj}
  \textbf{\bibinfo{volume}{598}}, \bibinfo{pages}{49} (\bibinfo{year}{2003}),
  \eprint{arXiv:astro-ph/0304292}.

\bibitem[{\citenamefont{{Peter} et~al.}(2010)\citenamefont{{Peter}, {Moody},
  and {Kamionkowski}}}]{peter_etal10}
\bibinfo{author}{\bibfnamefont{A.~H.~G.} \bibnamefont{{Peter}}},
  \bibinfo{author}{\bibfnamefont{C.~E.} \bibnamefont{{Moody}}},
  \bibnamefont{and}
  \bibinfo{author}{\bibfnamefont{M.}~\bibnamefont{{Kamionkowski}}},
  \bibinfo{journal}{ArXiv e-prints}  (\bibinfo{year}{2010}),
  \eprint{arXiv:1003.0419}.

\bibitem[{\citenamefont{{Spergel} and {Steinhardt}}(2000)}]{Spergel_etal00}
\bibinfo{author}{\bibfnamefont{D.~N.} \bibnamefont{{Spergel}}}
  \bibnamefont{and} \bibinfo{author}{\bibfnamefont{P.~J.}
  \bibnamefont{{Steinhardt}}}, \bibinfo{journal}{Physical Review Letters}
  \textbf{\bibinfo{volume}{84}}, \bibinfo{pages}{3760} (\bibinfo{year}{2000}),
  \eprint{arXiv:astro-ph/9909386}.

\bibitem[{\citenamefont{{Peter}}(2010{\natexlab{a}})}]{peter_10}
\bibinfo{author}{\bibfnamefont{A.~H.~G.} \bibnamefont{{Peter}}},
  \bibinfo{journal}{\prd} \textbf{\bibinfo{volume}{81}},
  \bibinfo{pages}{083511} (\bibinfo{year}{2010}{\natexlab{a}}),
  \eprint{arXiv:1001.3870}.

\bibitem[{\citenamefont{{S{\'a}nchez-Salcedo}}(2003)}]{Sanchez-Salcedo_03}
\bibinfo{author}{\bibfnamefont{F.~J.} \bibnamefont{{S{\'a}nchez-Salcedo}}},
  \bibinfo{journal}{The Astrophysical Journal Letters}
  \textbf{\bibinfo{volume}{591}}, \bibinfo{pages}{L107} (\bibinfo{year}{2003}),
  \eprint{arXiv:astro-ph/0305496}.

\bibitem[{\citenamefont{{Cen}}(2001)}]{cen_01}
\bibinfo{author}{\bibfnamefont{R.}~\bibnamefont{{Cen}}},
  \bibinfo{journal}{\apjl} \textbf{\bibinfo{volume}{546}}, \bibinfo{pages}{L77}
  (\bibinfo{year}{2001}), \eprint{arXiv:astro-ph/0005206}.

\bibitem[{\citenamefont{{Kaplinghat}}(2005)}]{Kaplinghat_05}
\bibinfo{author}{\bibfnamefont{M.}~\bibnamefont{{Kaplinghat}}},
  \bibinfo{journal}{\prd} \textbf{\bibinfo{volume}{72}},
  \bibinfo{pages}{063510} (\bibinfo{year}{2005}),
  \eprint{arXiv:astro-ph/0507300}.

\bibitem[{\citenamefont{{Abdelqader} and {Melia}}(2008)}]{Abdelqader_etal08}
\bibinfo{author}{\bibfnamefont{M.}~\bibnamefont{{Abdelqader}}}
  \bibnamefont{and} \bibinfo{author}{\bibfnamefont{F.}~\bibnamefont{{Melia}}},
  \bibinfo{journal}{\mnras} \textbf{\bibinfo{volume}{388}},
  \bibinfo{pages}{1869} (\bibinfo{year}{2008}), \eprint{0806.0602}.

\bibitem[{\citenamefont{{Peter} and {Benson}}(2010)}]{peter_benson10}
\bibinfo{author}{\bibfnamefont{A.~H.~G.} \bibnamefont{{Peter}}}
  \bibnamefont{and} \bibinfo{author}{\bibfnamefont{A.~J.}
  \bibnamefont{{Benson}}}, \bibinfo{journal}{\prd}
  \textbf{\bibinfo{volume}{82}}, \bibinfo{pages}{123521}
  (\bibinfo{year}{2010}), \eprint{1009.1912}.

\bibitem[{\citenamefont{{Bell} et~al.}(2010)\citenamefont{{Bell}, {Galea}, and
  {Petraki}}}]{Bell_etal10}
\bibinfo{author}{\bibfnamefont{N.~F.} \bibnamefont{{Bell}}},
  \bibinfo{author}{\bibfnamefont{A.~J.} \bibnamefont{{Galea}}},
  \bibnamefont{and}
  \bibinfo{author}{\bibfnamefont{K.}~\bibnamefont{{Petraki}}},
  \bibinfo{journal}{\prd} \textbf{\bibinfo{volume}{82}},
  \bibinfo{pages}{023514} (\bibinfo{year}{2010}), \eprint{arXiv:1004.1008}.

\bibitem[{\citenamefont{{Bell} et~al.}(2011)\citenamefont{{Bell}, {Galea}, and
  {Volkas}}}]{Bell_etal11}
\bibinfo{author}{\bibfnamefont{N.~F.} \bibnamefont{{Bell}}},
  \bibinfo{author}{\bibfnamefont{A.~J.} \bibnamefont{{Galea}}},
  \bibnamefont{and} \bibinfo{author}{\bibfnamefont{R.~R.}
  \bibnamefont{{Volkas}}}, \bibinfo{journal}{\prd}
  \textbf{\bibinfo{volume}{83}}, \bibinfo{pages}{063504}
  (\bibinfo{year}{2011}), \eprint{arXiv:1012.0067}.

\bibitem[{\citenamefont{{Aoyama} et~al.}(2011)\citenamefont{{Aoyama}, {Ichiki},
  {Nitta}, and {Sugiyama}}}]{Aoyama_etal11}
\bibinfo{author}{\bibfnamefont{S.}~\bibnamefont{{Aoyama}}},
  \bibinfo{author}{\bibfnamefont{K.}~\bibnamefont{{Ichiki}}},
  \bibinfo{author}{\bibfnamefont{D.}~\bibnamefont{{Nitta}}}, \bibnamefont{and}
  \bibinfo{author}{\bibfnamefont{N.}~\bibnamefont{{Sugiyama}}},
  \bibinfo{journal}{ArXiv e-prints}  (\bibinfo{year}{2011}),
  \eprint{arXiv:1106.1984}.

\bibitem[{\citenamefont{{Wang} and {Zentner}}(2010)}]{wang_zentner10}
\bibinfo{author}{\bibfnamefont{M.-Y.} \bibnamefont{{Wang}}} \bibnamefont{and}
  \bibinfo{author}{\bibfnamefont{A.~R.} \bibnamefont{{Zentner}}},
  \bibinfo{journal}{\prd} \textbf{\bibinfo{volume}{82}}, \bibinfo{eid}{123507}
  (\bibinfo{year}{2010}), \eprint{1011.2774}.

\bibitem[{\citenamefont{Collaborations}(2009)}]{lsstbook}
\bibinfo{author}{\bibfnamefont{L.~S.} \bibnamefont{Collaborations}},
  \bibinfo{journal}{arXiv:0912.0201}  (\bibinfo{year}{2009}).

\bibitem[{\citenamefont{{Refregier} et~al.}(2010)\citenamefont{{Refregier},
  {Amara}, {Kitching}, {Rassat}, {Scaramella}, {Weller}, and {Euclid Imaging
  Consortium}}}]{eicbook}
\bibinfo{author}{\bibfnamefont{A.}~\bibnamefont{{Refregier}}},
  \bibinfo{author}{\bibfnamefont{A.}~\bibnamefont{{Amara}}},
  \bibinfo{author}{\bibfnamefont{T.~D.} \bibnamefont{{Kitching}}},
  \bibinfo{author}{\bibfnamefont{A.}~\bibnamefont{{Rassat}}},
  \bibinfo{author}{\bibfnamefont{R.}~\bibnamefont{{Scaramella}}},
  \bibinfo{author}{\bibfnamefont{J.}~\bibnamefont{{Weller}}}, \bibnamefont{and}
  \bibinfo{author}{\bibfnamefont{f.~t.} \bibnamefont{{Euclid Imaging
  Consortium}}}, \bibinfo{journal}{ArXiv:1001.0061}  (\bibinfo{year}{2010}).

\bibitem[{\citenamefont{{Reid} et~al.}(2010)\citenamefont{{Reid}, {Percival},
  {Eisenstein}, {Verde}, {Spergel}, {Skibba}, {Bahcall}, {Budavari}, {Frieman},
  {Fukugita} et~al.}}]{Reid_etal10}
\bibinfo{author}{\bibfnamefont{B.~A.} \bibnamefont{{Reid}}},
  \bibinfo{author}{\bibfnamefont{W.~J.} \bibnamefont{{Percival}}},
  \bibinfo{author}{\bibfnamefont{D.~J.} \bibnamefont{{Eisenstein}}},
  \bibinfo{author}{\bibfnamefont{L.}~\bibnamefont{{Verde}}},
  \bibinfo{author}{\bibfnamefont{D.~N.} \bibnamefont{{Spergel}}},
  \bibinfo{author}{\bibfnamefont{R.~A.} \bibnamefont{{Skibba}}},
  \bibinfo{author}{\bibfnamefont{N.~A.} \bibnamefont{{Bahcall}}},
  \bibinfo{author}{\bibfnamefont{T.}~\bibnamefont{{Budavari}}},
  \bibinfo{author}{\bibfnamefont{J.~A.} \bibnamefont{{Frieman}}},
  \bibinfo{author}{\bibfnamefont{M.}~\bibnamefont{{Fukugita}}},
  \bibnamefont{et~al.}, \bibinfo{journal}{\mnras}
  \textbf{\bibinfo{volume}{404}}, \bibinfo{pages}{60} (\bibinfo{year}{2010}),
  \eprint{arXiv:0907.1659}.

\bibitem[{\citenamefont{{Seljak} et~al.}(2006)\citenamefont{{Seljak},
  {Makarov}, {McDonald}, and {Trac}}}]{Seljak_etal06}
\bibinfo{author}{\bibfnamefont{U.}~\bibnamefont{{Seljak}}},
  \bibinfo{author}{\bibfnamefont{A.}~\bibnamefont{{Makarov}}},
  \bibinfo{author}{\bibfnamefont{P.}~\bibnamefont{{McDonald}}},
  \bibnamefont{and} \bibinfo{author}{\bibfnamefont{H.}~\bibnamefont{{Trac}}},
  \bibinfo{journal}{Physical Review Letters} \textbf{\bibinfo{volume}{97}},
  \bibinfo{eid}{191303} (\bibinfo{year}{2006}),
  \eprint{arXiv:astro-ph/0602430}.

\bibitem[{\citenamefont{{Boyarsky} et~al.}(2009)\citenamefont{{Boyarsky},
  {Lesgourgues}, {Ruchayskiy}, and {Viel}}}]{boyarsky_etal08}
\bibinfo{author}{\bibfnamefont{A.}~\bibnamefont{{Boyarsky}}},
  \bibinfo{author}{\bibfnamefont{J.}~\bibnamefont{{Lesgourgues}}},
  \bibinfo{author}{\bibfnamefont{O.}~\bibnamefont{{Ruchayskiy}}},
  \bibnamefont{and} \bibinfo{author}{\bibfnamefont{M.}~\bibnamefont{{Viel}}},
  \bibinfo{journal}{Journal of Cosmology and Astro-Particle Physics}
  \textbf{\bibinfo{volume}{5}}, \bibinfo{pages}{12} (\bibinfo{year}{2009}),
  \eprint{arXiv:0812.0010}.

\bibitem[{\citenamefont{{Viel} et~al.}(2010)\citenamefont{{Viel}, {Haehnelt},
  and {Springel}}}]{Viel_etal10}
\bibinfo{author}{\bibfnamefont{M.}~\bibnamefont{{Viel}}},
  \bibinfo{author}{\bibfnamefont{M.~G.} \bibnamefont{{Haehnelt}}},
  \bibnamefont{and}
  \bibinfo{author}{\bibfnamefont{V.}~\bibnamefont{{Springel}}},
  \bibinfo{journal}{\jcap} \textbf{\bibinfo{volume}{6}}, \bibinfo{pages}{15}
  (\bibinfo{year}{2010}), \eprint{arXiv:1003.2422}.

\bibitem[{\citenamefont{{Hannestad} et~al.}(2006)\citenamefont{{Hannestad},
  {Tu}, and {Wong}}}]{Hannestad_etal06}
\bibinfo{author}{\bibfnamefont{S.}~\bibnamefont{{Hannestad}}},
  \bibinfo{author}{\bibfnamefont{H.}~\bibnamefont{{Tu}}}, \bibnamefont{and}
  \bibinfo{author}{\bibfnamefont{Y.~Y.} \bibnamefont{{Wong}}},
  \bibinfo{journal}{Journal of Cosmology and Astro-Particle Physics}
  \textbf{\bibinfo{volume}{6}}, \bibinfo{pages}{25} (\bibinfo{year}{2006}),
  \eprint{arXiv:astro-ph/0603019}.

\bibitem[{\citenamefont{{Ichiki} et~al.}(2009)\citenamefont{{Ichiki}, {Takada},
  and {Takahashi}}}]{Ichiki_etal09}
\bibinfo{author}{\bibfnamefont{K.}~\bibnamefont{{Ichiki}}},
  \bibinfo{author}{\bibfnamefont{M.}~\bibnamefont{{Takada}}}, \bibnamefont{and}
  \bibinfo{author}{\bibfnamefont{T.}~\bibnamefont{{Takahashi}}},
  \bibinfo{journal}{\prd} \textbf{\bibinfo{volume}{79}},
  \bibinfo{pages}{023520} (\bibinfo{year}{2009}), \eprint{arXiv:0810.4921}.

\bibitem[{\citenamefont{{Markovic} et~al.}(2011)\citenamefont{{Markovic},
  {Bridle}, {Slosar}, and {Weller}}}]{Markovic_etal11}
\bibinfo{author}{\bibfnamefont{K.}~\bibnamefont{{Markovic}}},
  \bibinfo{author}{\bibfnamefont{S.}~\bibnamefont{{Bridle}}},
  \bibinfo{author}{\bibfnamefont{A.}~\bibnamefont{{Slosar}}}, \bibnamefont{and}
  \bibinfo{author}{\bibfnamefont{J.}~\bibnamefont{{Weller}}},
  \bibinfo{journal}{\jcap} \textbf{\bibinfo{volume}{1}}, \bibinfo{pages}{22}
  (\bibinfo{year}{2011}), \eprint{arXiv:1009.0218}.

\bibitem[{\citenamefont{{Albrecht} et~al.}(2006)\citenamefont{{Albrecht},
  {Bernstein}, {Cahn}, {Freedman}, {Hewitt}, {Hu}, {Huth}, {Kamionkowski},
  {Kolb}, {Knox} et~al.}}]{detf}
\bibinfo{author}{\bibfnamefont{A.}~\bibnamefont{{Albrecht}}},
  \bibinfo{author}{\bibfnamefont{G.}~\bibnamefont{{Bernstein}}},
  \bibinfo{author}{\bibfnamefont{R.}~\bibnamefont{{Cahn}}},
  \bibinfo{author}{\bibfnamefont{W.~L.} \bibnamefont{{Freedman}}},
  \bibinfo{author}{\bibfnamefont{J.}~\bibnamefont{{Hewitt}}},
  \bibinfo{author}{\bibfnamefont{W.}~\bibnamefont{{Hu}}},
  \bibinfo{author}{\bibfnamefont{J.}~\bibnamefont{{Huth}}},
  \bibinfo{author}{\bibfnamefont{M.}~\bibnamefont{{Kamionkowski}}},
  \bibinfo{author}{\bibfnamefont{E.~W.} \bibnamefont{{Kolb}}},
  \bibinfo{author}{\bibfnamefont{L.}~\bibnamefont{{Knox}}},
  \bibnamefont{et~al.}, \bibinfo{journal}{ArXiv Astrophysics e-prints}
  (\bibinfo{year}{2006}), \eprint{astro-ph/0609591}.

\bibitem[{\citenamefont{{Huterer}}(2010)}]{huterer10}
\bibinfo{author}{\bibfnamefont{D.}~\bibnamefont{{Huterer}}},
  \bibinfo{journal}{arXiv:1001.1758}  (\bibinfo{year}{2010}).

\bibitem[{\citenamefont{{Zentner} et~al.}(2008)\citenamefont{{Zentner}, {Rudd},
  and {Hu}}}]{zentner_etal08}
\bibinfo{author}{\bibfnamefont{A.~R.} \bibnamefont{{Zentner}}},
  \bibinfo{author}{\bibfnamefont{D.~H.} \bibnamefont{{Rudd}}},
  \bibnamefont{and} \bibinfo{author}{\bibfnamefont{W.}~\bibnamefont{{Hu}}},
  \bibinfo{journal}{\prd} \textbf{\bibinfo{volume}{77}},
  \bibinfo{pages}{043507} (\bibinfo{year}{2008}), \eprint{arXiv:0709.4029}.

\bibitem[{\citenamefont{{Ma} et~al.}(2006)\citenamefont{{Ma}, {Hu}, and
  {Huterer}}}]{ma_etal06}
\bibinfo{author}{\bibfnamefont{Z.}~\bibnamefont{{Ma}}},
  \bibinfo{author}{\bibfnamefont{W.}~\bibnamefont{{Hu}}}, \bibnamefont{and}
  \bibinfo{author}{\bibfnamefont{D.}~\bibnamefont{{Huterer}}},
  \bibinfo{journal}{\apj} \textbf{\bibinfo{volume}{636}}, \bibinfo{pages}{21}
  (\bibinfo{year}{2006}), \eprint{astro-ph/0506614}.

\bibitem[{\citenamefont{{Ma} and
  {Bertschinger}}(1995{\natexlab{a}})}]{ma_bertschinger95}
\bibinfo{author}{\bibfnamefont{C.}~\bibnamefont{{Ma}}} \bibnamefont{and}
  \bibinfo{author}{\bibfnamefont{E.}~\bibnamefont{{Bertschinger}}},
  \bibinfo{journal}{\apj} \textbf{\bibinfo{volume}{455}}, \bibinfo{pages}{7}
  (\bibinfo{year}{1995}{\natexlab{a}}), \eprint{arXiv:astro-ph/9506072}.

\bibitem[{\citenamefont{{Hearin} et~al.}(2010)\citenamefont{{Hearin},
  {Zentner}, {Ma}, and {Huterer}}}]{hearin_etal10}
\bibinfo{author}{\bibfnamefont{A.~P.} \bibnamefont{{Hearin}}},
  \bibinfo{author}{\bibfnamefont{A.~R.} \bibnamefont{{Zentner}}},
  \bibinfo{author}{\bibfnamefont{Z.}~\bibnamefont{{Ma}}}, \bibnamefont{and}
  \bibinfo{author}{\bibfnamefont{D.}~\bibnamefont{{Huterer}}},
  \bibinfo{journal}{\apj} \textbf{\bibinfo{volume}{720}}, \bibinfo{pages}{1351}
  (\bibinfo{year}{2010}), \eprint{arXiv:1002.3383}.

\bibitem[{\citenamefont{{Smail}
  et~al.}(1995{\natexlab{a}})\citenamefont{{Smail}, {Hogg}, {Yan}, and
  {Cohen}}}]{smail_etal95a}
\bibinfo{author}{\bibfnamefont{I.}~\bibnamefont{{Smail}}},
  \bibinfo{author}{\bibfnamefont{D.~W.} \bibnamefont{{Hogg}}},
  \bibinfo{author}{\bibfnamefont{L.}~\bibnamefont{{Yan}}}, \bibnamefont{and}
  \bibinfo{author}{\bibfnamefont{J.~G.} \bibnamefont{{Cohen}}},
  \bibinfo{journal}{\apjl} \textbf{\bibinfo{volume}{449}},
  \bibinfo{pages}{L105+} (\bibinfo{year}{1995}{\natexlab{a}}),
  \eprint{arXiv:astro-ph/9506095}.

\bibitem[{\citenamefont{{Smail}
  et~al.}(1995{\natexlab{b}})\citenamefont{{Smail}, {Ellis}, {Fitchett}, and
  {Edge}}}]{smail_etal95b}
\bibinfo{author}{\bibfnamefont{I.}~\bibnamefont{{Smail}}},
  \bibinfo{author}{\bibfnamefont{R.~S.} \bibnamefont{{Ellis}}},
  \bibinfo{author}{\bibfnamefont{M.~J.} \bibnamefont{{Fitchett}}},
  \bibnamefont{and} \bibinfo{author}{\bibfnamefont{A.~C.}
  \bibnamefont{{Edge}}}, \bibinfo{journal}{\mnras}
  \textbf{\bibinfo{volume}{273}}, \bibinfo{pages}{277}
  (\bibinfo{year}{1995}{\natexlab{b}}), \eprint{arXiv:astro-ph/9402049}.

\bibitem[{\citenamefont{{Newman}}(2008)}]{newman08}
\bibinfo{author}{\bibfnamefont{J.~A.} \bibnamefont{{Newman}}},
  \bibinfo{journal}{\apj} \textbf{\bibinfo{volume}{684}}, \bibinfo{pages}{88}
  (\bibinfo{year}{2008}), \eprint{arXiv:0805.1409}.

\bibitem[{\citenamefont{{Bernstein} and {Huterer}}(2010)}]{bernstein_huterer10}
\bibinfo{author}{\bibfnamefont{G.}~\bibnamefont{{Bernstein}}} \bibnamefont{and}
  \bibinfo{author}{\bibfnamefont{D.}~\bibnamefont{{Huterer}}},
  \bibinfo{journal}{\mnras} \textbf{\bibinfo{volume}{401}},
  \bibinfo{pages}{1399} (\bibinfo{year}{2010}), \eprint{arXiv:0902.2782}.

\bibitem[{\citenamefont{{Massey} et~al.}(2004)\citenamefont{{Massey}, {Rhodes},
  {Refregier}, {Albert}, {Bacon}, {Bernstein}, {Ellis}, {Jain}, {McKay},
  {Perlmutter} et~al.}}]{massey_etal04}
\bibinfo{author}{\bibfnamefont{R.}~\bibnamefont{{Massey}}},
  \bibinfo{author}{\bibfnamefont{J.}~\bibnamefont{{Rhodes}}},
  \bibinfo{author}{\bibfnamefont{A.}~\bibnamefont{{Refregier}}},
  \bibinfo{author}{\bibfnamefont{J.}~\bibnamefont{{Albert}}},
  \bibinfo{author}{\bibfnamefont{D.}~\bibnamefont{{Bacon}}},
  \bibinfo{author}{\bibfnamefont{G.}~\bibnamefont{{Bernstein}}},
  \bibinfo{author}{\bibfnamefont{R.}~\bibnamefont{{Ellis}}},
  \bibinfo{author}{\bibfnamefont{B.}~\bibnamefont{{Jain}}},
  \bibinfo{author}{\bibfnamefont{T.}~\bibnamefont{{McKay}}},
  \bibinfo{author}{\bibfnamefont{S.}~\bibnamefont{{Perlmutter}}},
  \bibnamefont{et~al.}, \bibinfo{journal}{\aj} \textbf{\bibinfo{volume}{127}},
  \bibinfo{pages}{3089} (\bibinfo{year}{2004}),
  \eprint{arXiv:astro-ph/0304418}.

\bibitem[{\citenamefont{{Kasliwal} et~al.}(2008)\citenamefont{{Kasliwal},
  {Massey}, {Ellis}, {Miyazaki}, and {Rhodes}}}]{kasliwal_etal08}
\bibinfo{author}{\bibfnamefont{M.~M.} \bibnamefont{{Kasliwal}}},
  \bibinfo{author}{\bibfnamefont{R.}~\bibnamefont{{Massey}}},
  \bibinfo{author}{\bibfnamefont{R.~S.} \bibnamefont{{Ellis}}},
  \bibinfo{author}{\bibfnamefont{S.}~\bibnamefont{{Miyazaki}}},
  \bibnamefont{and} \bibinfo{author}{\bibfnamefont{J.}~\bibnamefont{{Rhodes}}},
  \bibinfo{journal}{\apj} \textbf{\bibinfo{volume}{684}}, \bibinfo{pages}{34}
  (\bibinfo{year}{2008}), \eprint{0710.3588}.

\bibitem[{\citenamefont{{White} and {Hu}}(2000)}]{white_hu00}
\bibinfo{author}{\bibfnamefont{M.}~\bibnamefont{{White}}} \bibnamefont{and}
  \bibinfo{author}{\bibfnamefont{W.}~\bibnamefont{{Hu}}},
  \bibinfo{journal}{\apj} \textbf{\bibinfo{volume}{537}}, \bibinfo{pages}{1}
  (\bibinfo{year}{2000}).

\bibitem[{\citenamefont{{Cooray} and {Hu}}(2001)}]{cooray_hu01}
\bibinfo{author}{\bibfnamefont{A.}~\bibnamefont{{Cooray}}} \bibnamefont{and}
  \bibinfo{author}{\bibfnamefont{W.}~\bibnamefont{{Hu}}},
  \bibinfo{journal}{\apj} \textbf{\bibinfo{volume}{554}}, \bibinfo{pages}{56}
  (\bibinfo{year}{2001}), \eprint{astro-ph/0012087}.

\bibitem[{\citenamefont{{Vale} and {White}}(2003)}]{vale_white03}
\bibinfo{author}{\bibfnamefont{C.}~\bibnamefont{{Vale}}} \bibnamefont{and}
  \bibinfo{author}{\bibfnamefont{M.}~\bibnamefont{{White}}},
  \bibinfo{journal}{Apj} \textbf{\bibinfo{volume}{592}}, \bibinfo{pages}{699}
  (\bibinfo{year}{2003}).

\bibitem[{\citenamefont{{Dodelson} et~al.}(2006)\citenamefont{{Dodelson},
  {Shapiro}, and {White}}}]{dodelson_etal06}
\bibinfo{author}{\bibfnamefont{S.}~\bibnamefont{{Dodelson}}},
  \bibinfo{author}{\bibfnamefont{C.}~\bibnamefont{{Shapiro}}},
  \bibnamefont{and} \bibinfo{author}{\bibfnamefont{M.}~\bibnamefont{{White}}},
  \bibinfo{journal}{\prd} \textbf{\bibinfo{volume}{73}},
  \bibinfo{pages}{023009} (\bibinfo{year}{2006}),
  \eprint{arXiv:astro-ph/0508296}.

\bibitem[{\citenamefont{{Semboloni} et~al.}(2007)\citenamefont{{Semboloni},
  {van Waerbeke}, {Heymans}, {Hamana}, {Colombi}, {White}, and
  {Mellier}}}]{semboloni_etal06}
\bibinfo{author}{\bibfnamefont{E.}~\bibnamefont{{Semboloni}}},
  \bibinfo{author}{\bibfnamefont{L.}~\bibnamefont{{van Waerbeke}}},
  \bibinfo{author}{\bibfnamefont{C.}~\bibnamefont{{Heymans}}},
  \bibinfo{author}{\bibfnamefont{T.}~\bibnamefont{{Hamana}}},
  \bibinfo{author}{\bibfnamefont{S.}~\bibnamefont{{Colombi}}},
  \bibinfo{author}{\bibfnamefont{M.}~\bibnamefont{{White}}}, \bibnamefont{and}
  \bibinfo{author}{\bibfnamefont{Y.}~\bibnamefont{{Mellier}}},
  \bibinfo{journal}{\mnras} \textbf{\bibinfo{volume}{375}}, \bibinfo{pages}{L6}
  (\bibinfo{year}{2007}), \eprint{arXiv:astro-ph/0606648}.

\bibitem[{\citenamefont{{Starkman} et~al.}(1994)\citenamefont{{Starkman},
  {Kaiser}, and {Malaney}}}]{Starkman_etal94}
\bibinfo{author}{\bibfnamefont{G.~D.} \bibnamefont{{Starkman}}},
  \bibinfo{author}{\bibfnamefont{N.}~\bibnamefont{{Kaiser}}}, \bibnamefont{and}
  \bibinfo{author}{\bibfnamefont{R.~A.} \bibnamefont{{Malaney}}},
  \bibinfo{journal}{\apj} \textbf{\bibinfo{volume}{434}}, \bibinfo{pages}{12}
  (\bibinfo{year}{1994}), \eprint{arXiv:astro-ph/9312020}.

\bibitem[{\citenamefont{{Lopez}}(1999)}]{Lopez_etal99}
\bibinfo{author}{\bibfnamefont{R.~E.} \bibnamefont{{Lopez}}}, Ph.D. thesis,
  \bibinfo{school}{THE UNIVERSITY OF CHICAGO} (\bibinfo{year}{1999}).

\bibitem[{\citenamefont{{Kolb} and {Turner}}(1994)}]{Kolb_Turner}
\bibinfo{author}{\bibfnamefont{E.}~\bibnamefont{{Kolb}}} \bibnamefont{and}
  \bibinfo{author}{\bibfnamefont{M.}~\bibnamefont{{Turner}}},
  \emph{\bibinfo{title}{{The Early Univrse}}} (\bibinfo{publisher}{{Westview
  Press}}, \bibinfo{year}{1994}).

\bibitem[{\citenamefont{{Ma} and
  {Bertschinger}}(1995{\natexlab{b}})}]{Ma_etal95}
\bibinfo{author}{\bibfnamefont{C.}~\bibnamefont{{Ma}}} \bibnamefont{and}
  \bibinfo{author}{\bibfnamefont{E.}~\bibnamefont{{Bertschinger}}},
  \bibinfo{journal}{\apj} \textbf{\bibinfo{volume}{455}}, \bibinfo{pages}{7}
  (\bibinfo{year}{1995}{\natexlab{b}}), \eprint{arXiv:astro-ph/9506072}.

\bibitem[{\citenamefont{{Peter}}(2010{\natexlab{b}})}]{peter10}
\bibinfo{author}{\bibfnamefont{A.~H.~G.} \bibnamefont{{Peter}}},
  \bibinfo{journal}{ArXiv e-prints}  (\bibinfo{year}{2010}{\natexlab{b}}),
  \eprint{arXiv:1001.3870}.

\bibitem[{\citenamefont{{Heitmann} et~al.}(2008)\citenamefont{{Heitmann},
  {White}, {Wagner}, {Habib}, and {Higdon}}}]{heitmann_etal08}
\bibinfo{author}{\bibfnamefont{K.}~\bibnamefont{{Heitmann}}},
  \bibinfo{author}{\bibfnamefont{M.}~\bibnamefont{{White}}},
  \bibinfo{author}{\bibfnamefont{C.}~\bibnamefont{{Wagner}}},
  \bibinfo{author}{\bibfnamefont{S.}~\bibnamefont{{Habib}}}, \bibnamefont{and}
  \bibinfo{author}{\bibfnamefont{D.}~\bibnamefont{{Higdon}}},
  \bibinfo{journal}{ArXiv:0812.1052}  (\bibinfo{year}{2008}).

\bibitem[{\citenamefont{{Cooray} and {Sheth}}(2002)}]{cooray_sheth02}
\bibinfo{author}{\bibfnamefont{A.}~\bibnamefont{{Cooray}}} \bibnamefont{and}
  \bibinfo{author}{\bibfnamefont{R.}~\bibnamefont{{Sheth}}},
  \bibinfo{journal}{\physrep} \textbf{\bibinfo{volume}{372}},
  \bibinfo{pages}{1} (\bibinfo{year}{2002}).

\bibitem[{\citenamefont{{Smith} et~al.}(2003)\citenamefont{{Smith}, {Peacock},
  {Jenkins}, {White}, {Frenk}, {Pearce}, {Thomas}, {Efstathiou}, and
  {Couchman}}}]{smith_etal03}
\bibinfo{author}{\bibfnamefont{R.~E.} \bibnamefont{{Smith}}},
  \bibinfo{author}{\bibfnamefont{J.~A.} \bibnamefont{{Peacock}}},
  \bibinfo{author}{\bibfnamefont{A.}~\bibnamefont{{Jenkins}}},
  \bibinfo{author}{\bibfnamefont{S.~D.~M.} \bibnamefont{{White}}},
  \bibinfo{author}{\bibfnamefont{C.~S.} \bibnamefont{{Frenk}}},
  \bibinfo{author}{\bibfnamefont{F.~R.} \bibnamefont{{Pearce}}},
  \bibinfo{author}{\bibfnamefont{P.~A.} \bibnamefont{{Thomas}}},
  \bibinfo{author}{\bibfnamefont{G.}~\bibnamefont{{Efstathiou}}},
  \bibnamefont{and} \bibinfo{author}{\bibfnamefont{H.~M.~P.}
  \bibnamefont{{Couchman}}}, \bibinfo{journal}{\mnras}
  \textbf{\bibinfo{volume}{341}}, \bibinfo{pages}{1311} (\bibinfo{year}{2003}),
  \eprint{astro-ph/0207664}.

\bibitem[{\citenamefont{{Navarro} et~al.}(1997)\citenamefont{{Navarro},
  {Frenk}, and {White}}}]{navarro_etal97}
\bibinfo{author}{\bibfnamefont{J.~F.} \bibnamefont{{Navarro}}},
  \bibinfo{author}{\bibfnamefont{C.~S.} \bibnamefont{{Frenk}}},
  \bibnamefont{and} \bibinfo{author}{\bibfnamefont{S.~D.~M.}
  \bibnamefont{{White}}}, \bibinfo{journal}{\apj}
  \textbf{\bibinfo{volume}{490}}, \bibinfo{pages}{493} (\bibinfo{year}{1997}),
  \eprint{astro-ph/9611107}.

\bibitem[{\citenamefont{{Jungman}
  et~al.}(1996{\natexlab{b}})\citenamefont{{Jungman}, {Kamionkowski},
  {Kosowsky}, and {Spergel}}}]{jungman_etal96}
\bibinfo{author}{\bibfnamefont{G.}~\bibnamefont{{Jungman}}},
  \bibinfo{author}{\bibfnamefont{M.}~\bibnamefont{{Kamionkowski}}},
  \bibinfo{author}{\bibfnamefont{A.}~\bibnamefont{{Kosowsky}}},
  \bibnamefont{and} \bibinfo{author}{\bibfnamefont{D.~N.}
  \bibnamefont{{Spergel}}}, \bibinfo{journal}{\prd}
  \textbf{\bibinfo{volume}{54}}, \bibinfo{pages}{1332}
  (\bibinfo{year}{1996}{\natexlab{b}}), \eprint{arXiv:astro-ph/9512139}.

\bibitem[{\citenamefont{{Tegmark} et~al.}(1997)\citenamefont{{Tegmark},
  {Taylor}, and {Heavens}}}]{tegmark_etal97}
\bibinfo{author}{\bibfnamefont{M.}~\bibnamefont{{Tegmark}}},
  \bibinfo{author}{\bibfnamefont{A.~N.} \bibnamefont{{Taylor}}},
  \bibnamefont{and} \bibinfo{author}{\bibfnamefont{A.~F.}
  \bibnamefont{{Heavens}}}, \bibinfo{journal}{\apj}
  \textbf{\bibinfo{volume}{480}}, \bibinfo{pages}{22} (\bibinfo{year}{1997}),
  \eprint{arXiv:astro-ph/9603021}.

\bibitem[{\citenamefont{{Seljak}}(1997)}]{seljak97}
\bibinfo{author}{\bibfnamefont{U.}~\bibnamefont{{Seljak}}},
  \bibinfo{journal}{\apj} \textbf{\bibinfo{volume}{482}}, \bibinfo{pages}{6}
  (\bibinfo{year}{1997}), \eprint{arXiv:astro-ph/9608131}.

\bibitem[{\citenamefont{{Hu}}(1999)}]{hu99}
\bibinfo{author}{\bibfnamefont{W.}~\bibnamefont{{Hu}}},
  \bibinfo{journal}{\apjl} \textbf{\bibinfo{volume}{522}}, \bibinfo{pages}{L21}
  (\bibinfo{year}{1999}), \eprint{astro-ph/9904153}.

\bibitem[{\citenamefont{{Kosowsky} et~al.}(2002)\citenamefont{{Kosowsky},
  {Milosavljevic}, and {Jimenez}}}]{kosowsky_etal02}
\bibinfo{author}{\bibfnamefont{A.}~\bibnamefont{{Kosowsky}}},
  \bibinfo{author}{\bibfnamefont{M.}~\bibnamefont{{Milosavljevic}}},
  \bibnamefont{and}
  \bibinfo{author}{\bibfnamefont{R.}~\bibnamefont{{Jimenez}}},
  \bibinfo{journal}{\prd} \textbf{\bibinfo{volume}{66}},
  \bibinfo{pages}{063007} (\bibinfo{year}{2002}),
  \eprint{arXiv:astro-ph/0206014}.

\bibitem[{\citenamefont{{Huterer} and {Takada}}(2005)}]{huterer_takada05}
\bibinfo{author}{\bibfnamefont{D.}~\bibnamefont{{Huterer}}} \bibnamefont{and}
  \bibinfo{author}{\bibfnamefont{M.}~\bibnamefont{{Takada}}},
  \bibinfo{journal}{Astroparticle Physics} \textbf{\bibinfo{volume}{23}},
  \bibinfo{pages}{369} (\bibinfo{year}{2005}), \eprint{astro-ph/0412142}.

\bibitem[{\citenamefont{{Kitching} et~al.}(2008)\citenamefont{{Kitching},
  {Heavens}, {Verde}, {Serra}, and {Melchiorri}}}]{kitching_etal08}
\bibinfo{author}{\bibfnamefont{T.~D.} \bibnamefont{{Kitching}}},
  \bibinfo{author}{\bibfnamefont{A.~F.} \bibnamefont{{Heavens}}},
  \bibinfo{author}{\bibfnamefont{L.}~\bibnamefont{{Verde}}},
  \bibinfo{author}{\bibfnamefont{P.}~\bibnamefont{{Serra}}}, \bibnamefont{and}
  \bibinfo{author}{\bibfnamefont{A.}~\bibnamefont{{Melchiorri}}},
  \bibinfo{journal}{\prd} \textbf{\bibinfo{volume}{77}},
  \bibinfo{pages}{103008} (\bibinfo{year}{2008}), \eprint{arXiv:0801.4565}.

\bibitem[{\citenamefont{{Peter}}(2009)}]{peter09}
\bibinfo{author}{\bibfnamefont{A.~H.~G.} \bibnamefont{{Peter}}},
  \bibinfo{journal}{ArXiv e-prints}  (\bibinfo{year}{2009}),
  \eprint{arXiv:0910.4765}.

\bibitem[{\citenamefont{{Rudd} et~al.}(2008)\citenamefont{{Rudd}, {Zentner},
  and {Kravtsov}}}]{rudd_etal08}
\bibinfo{author}{\bibfnamefont{D.~H.} \bibnamefont{{Rudd}}},
  \bibinfo{author}{\bibfnamefont{A.~R.} \bibnamefont{{Zentner}}},
  \bibnamefont{and} \bibinfo{author}{\bibfnamefont{A.~V.}
  \bibnamefont{{Kravtsov}}}, \bibinfo{journal}{\apj}
  \textbf{\bibinfo{volume}{672}}, \bibinfo{pages}{19} (\bibinfo{year}{2008}),
  \eprint{arXiv:astro-ph/0703741}.

\bibitem[{\citenamefont{{Hu} et~al.}(2006)\citenamefont{{Hu}, {Huterer}, and
  {Smith}}}]{hu_etal06}
\bibinfo{author}{\bibfnamefont{W.}~\bibnamefont{{Hu}}},
  \bibinfo{author}{\bibfnamefont{D.}~\bibnamefont{{Huterer}}},
  \bibnamefont{and} \bibinfo{author}{\bibfnamefont{K.~M.}
  \bibnamefont{{Smith}}}, \bibinfo{journal}{\apjl}
  \textbf{\bibinfo{volume}{650}}, \bibinfo{pages}{L13} (\bibinfo{year}{2006}),
  \eprint{arXiv:astro-ph/0607316}.

\bibitem[{\citenamefont{{Bird} et~al.}(2011)\citenamefont{{Bird}, {Viel}, and
  {Haehnelt}}}]{Bird_etal11}
\bibinfo{author}{\bibfnamefont{S.}~\bibnamefont{{Bird}}},
  \bibinfo{author}{\bibfnamefont{M.}~\bibnamefont{{Viel}}}, \bibnamefont{and}
  \bibinfo{author}{\bibfnamefont{M.~G.} \bibnamefont{{Haehnelt}}},
  \bibinfo{journal}{ArXiv e-prints}  (\bibinfo{year}{2011}),
  \eprint{arXiv:1109.4416}.

\bibitem[{\citenamefont{{Saito} et~al.}(2008)\citenamefont{{Saito}, {Takada},
  and {Taruya}}}]{Saito_etal08}
\bibinfo{author}{\bibfnamefont{S.}~\bibnamefont{{Saito}}},
  \bibinfo{author}{\bibfnamefont{M.}~\bibnamefont{{Takada}}}, \bibnamefont{and}
  \bibinfo{author}{\bibfnamefont{A.}~\bibnamefont{{Taruya}}},
  \bibinfo{journal}{Physical Review Letters} \textbf{\bibinfo{volume}{100}},
  \bibinfo{eid}{191301} (\bibinfo{year}{2008}), \eprint{0801.0607}.

\bibitem[{\citenamefont{{Zhan} and {Knox}}(2006)}]{zhan_knox06}
\bibinfo{author}{\bibfnamefont{H.}~\bibnamefont{{Zhan}}} \bibnamefont{and}
  \bibinfo{author}{\bibfnamefont{L.}~\bibnamefont{{Knox}}},
  \bibinfo{journal}{ArXiv Astrophysics e-prints}  (\bibinfo{year}{2006}),
  \eprint{arXiv:astro-ph/0611159}.

\bibitem[{\citenamefont{{Hearin} and {Zentner}}(2009)}]{hearin_zentner09}
\bibinfo{author}{\bibfnamefont{A.~P.} \bibnamefont{{Hearin}}} \bibnamefont{and}
  \bibinfo{author}{\bibfnamefont{A.~R.} \bibnamefont{{Zentner}}},
  \bibinfo{journal}{Journal of Cosmology and Astro-Particle Physics}
  \textbf{\bibinfo{volume}{4}}, \bibinfo{pages}{32} (\bibinfo{year}{2009}),
  \eprint{arXiv:0904.3334}.

\end{thebibliography}


\end{document}